\documentclass[journal,10pt]{IEEEtran}
\usepackage{amsmath,amsfonts,amssymb}
\usepackage{mathrsfs}
\usepackage[noend]{algpseudocode}
\usepackage{algorithmicx}
\usepackage{algorithm}
\usepackage{array}
\pdfoutput=1
\usepackage{textcomp}
\usepackage{stfloats}
\usepackage{url}
\usepackage{xcolor}
\usepackage{verbatim}
\usepackage{graphicx}
\usepackage{subcaption}
\usepackage{epstopdf}
\usepackage{cite}
\usepackage{makecell,multirow}
\usepackage{enumitem}
\def\BibTeX{{\rm B\kern-.05em{\sc i\kern-.025em b}\kern-.08em
    T\kern-.1667em\lower.7ex\hbox{E}\kern-.125emX}}

\begin{document}

\title{Maximum Likelihood CFO Estimation for High-Mobility OFDM Systems: A Chinese Remainder Theorem Based Method}

\author{Wei Huang, Jun Wang,~\IEEEmembership{Senior Member,~IEEE}, Xiaoping Li, Qihang Peng
\thanks{Wei Huang and Jun Wang are with the National Key Laboratory of Wireless Communications, University of Electronic Science and Technology of China, Chengdu, 611731, China (e-mail: 18581810911@163.com, junwang@uestc.edu.cn).}
\thanks{Xiaoping Li is with the School of Mathematical Science, University of Electronic Science and Technology of China, Chengdu, 611731, China (e-mail: lixiaoping.math@uestc.edu.cn).}
\thanks{Qihang Peng is with the School of Information and Communication Engineering, University of Electronic Science and Technology of China, Chengdu, 611731, China (e-mail: anniepqh@uestc.edu.cn).}}

\markboth{Journal of \LaTeX\ Class Files,~Vol.~14, No.~8, August~2023}%
{Shell \MakeLowercase{\textit{et al.}}: A Sample Article Using IEEEtran.cls for IEEE Journals}

\maketitle
\begin{abstract}
Orthogonal frequency division multiplexing (OFDM) is a widely adopted wireless communication technique but is sensitive to the carrier frequency offset (CFO).
For high-mobility environments, severe Doppler shifts cause the CFO to extend well beyond the subcarrier spacing.
Traditional algorithms generally estimate the integer and fractional parts of the CFO separately, which is time-consuming and requires high additional computations.
To address these issues, this paper proposes a Chinese remainder theorem-based CFO Maximum Likelihood Estimation (CCMLE) approach for jointly estimating the integer and fractional parts.
With CCMLE, the MLE of the CFO can be obtained directly from multiple estimates of sequences with varying lengths. This approach can achieve a wide estimation range up to the total number of subcarriers, without significant additional computations.
Furthermore, we show that the CCMLE can approach the Cram$\acute{\text{e}}$r-Rao Bound (CRB), and give an analytic expression for the signal-to-noise ratio (SNR) threshold approaching the CRB, enabling an efficient waveform design.
Accordingly, a parameter configuration guideline for the CCMLE is presented to achieve a better MSE performance and a lower SNR threshold.
Finally, experiments show that our proposed method is highly consistent with the theoretical analysis and advantageous regarding estimated range and error performance compared to baselines.
\end{abstract}


\section{Introduction}
\IEEEPARstart{S}{atellite} communication is expected to take center stage for the 6th generation of mobile communication systems (6G) commercial services due to its wide bandwidth, ubiquitous coverage, and ability to withstand natural disasters \cite{zhou2023aerospace,nguyen20216g}.
Meanwhile, orthogonal frequency division multiplexing (OFDM) is the primary scheme of an integrated satellite-terrestrial network, given its promising anti-multipath fading characteristics and high spectrum efficiency \cite{shafi20175g,dogra2020survey,huang2021synchronization}.
However, a crucial feature of the satellite network is that the high mobility leads to significant Doppler shifts, and an inherent drawback of OFDM systems is that they are very sensitive to the carrier frequency offset (CFO) \cite{pollet1995ber}, including CFOs at integer and fractional multiples of the subcarrier spacing.
Integer frequency offset (IFO) will result in a cyclic shift of the subcarriers while fractional frequency offset (FFO) will destroy the orthogonality of the subcarriers and lead to inter-carrier interference (ICI) \cite{cho2010mimo}.
Both kinds of CFOs can lead to severe performance degradation if not accurately compensated.
Therefore, an accurate estimation of the CFO is of critical importance in high-mobility OFDM systems, where Doppler shifts can be well beyond the subcarrier spacing.

In general, the frequency synchronization algorithms for OFDM systems can be classified into two categories, i.e., non-data-aided blind estimation methods \cite{bolcskei2001blind,van1997ml,chen2004blind} and data-aided estimation methods\cite{moose1994technique,bai2013comments,schmidl1997robust,kim2001efficient,gul2014timing,huang2021synchronization}.
Most of the non-data-aided blind estimation methods perform synchronization based on the inherent structure of the OFDM symbol and generally have high latency and computational complexity \cite{huang2021synchronization,gul2014timing}, which could not meet the fast and reliable requirements of many applications.
In contrast, data-aided frequency synchronization schemes typically use the autocorrelation of received training symbols or the cross-correlation with the local copy at the receiver \cite{gul2014timing}, which is faster, more reliable, and easier to implement.
Therefore, the data-aided frequency synchronization methods are more suitable for practical applications due to their low latency and high reliability \cite{huang2021synchronization}.

For the data-aided methods, early research works focused on the FFO estimation problem.
In \cite{moose1994technique,bai2013comments}, Moose and Bai presented a maximum likelihood estimation (MLE) of the FFO based on two repeated symbols.
Although the accuracy of the CFO estimate is satisfying, the estimated CFO range of these method is limited, i.e., $\pm1/2$ the subcarrier spacing.
In fact, the accuracy and range of the CFO estimate are generally in conflict with each other, e.g., a wider estimated range usually requires shorter sequences, which will deteriorate the accuracy \cite{huang2021synchronization}.
In order to increase the range of the CFO estimate while maintaining the accuracy of the algorithm, an effective alternative approach is to perform accurate FFO estimation over a small range and then perform IFO estimation based on the FFO-compensated signal \cite{schmidl1997robust,kim2001efficient,gul2014timing,huang2021synchronization}.
In \cite{schmidl1997robust}, Schmidl presented a wide-range frequency synchronization based on the autocorrelation of two training OFDM symbols.
Further, authors in \cite{kim2001efficient} proposed an improved estimator based on the timing synchronization algorithm in \cite{schmidl1997robust}, which can achieve the same CFO range as \cite{schmidl1997robust} but only one training symbol is required.
Considering the sensitivity to CFO for the Zadoff-Chu (ZC) sequences-based timing synchronization method, \cite{gul2014timing} proposed a cross-correlation-based joint time and frequency synchronization schemes for OFDM downlink transmissions using two suitable ZC sequences.
Furthermore, a frequency synchronization methods for the OFDM-based satellite system was proposed by fully utilizing the correlation property of two symmetric ZC sequences\cite{huang2021synchronization}.
However, these methods are time-consuming due to multi-level estimation and requires a large amount of additional computations to estimate IFO.
To address these issues, the Chinese remainder theorem (CRT) are applied \cite{rha2015real,huang2018doppler,wei2022sequential}, which can directly obtain estimation results with a full estimated range at the cost of a small number of additional computations.

The CRT is to reconstruct a number by the remainder of a series of integer modes \cite{wang2015maximum}, which is widely used in the field of sub-Nyquist-sampling frequency determination \cite{xiao2018frequency}, DOA estimation \cite{li2018robust}, range estimation \cite{chen2021rangeEstimation}, and physical layer encryption \cite{zhang2021encryption}.
As a first attempt, to obtain a wide range while retaining a high accuracy for the CFO estimation, authors in \cite{rha2015real} applied the classic CRT to estimate the integer parts of the CFO with two relatively co-prime sample intervals used in coherent optical OFDM systems.
However, this method could not directly obtain the fractional parts of the CFO estimation, but requires a length of the product of all sample interval values to estimate the FFO individually.
In order to jointly estimate the integer and fractional parts of a real number with some small sample intervals, the robust CRT \cite{wang2010closed} are considered.
Authors in \cite{huang2018doppler} removed the phase uncertainty for Doppler shift detection in synthetic aperture radar (SAR) systems based on the closed-form robust CRT, which allows for the direct CFO estimate of a real number.
Furthermore, \cite{wei2022sequential} also used a sequential-based approach to transform the Doppler estimation problem into a robust CRT problem and solved by the closed-form robust CRT.
However, the closed-form robust CRT are based on the assumption that all the remainder errors have the same variance \cite{wang2010closed}, i.e., the CFO estimation errors for different estimation ranges have the same variance.
In this case, a closed-form robust CRT actually can only achieve the suboptimal performance \cite{wang2015maximum}, since the errors usually vary differently over different estimation ranges.
Hence, the robust CRT-based method in \cite{huang2018doppler,wei2022sequential} for Doppler shift detection also yields only the suboptimal performance for the CFO estimation problem.\\
\indent Considering the variability in the distribution of CFO estimation errors obtained under different estimated ranges, we propose a CRT-based CFO Maximum Likelihood Estimation (CCMLE) method, which enables MLE of CFOs based on multiple CFO estimates from sample intervals of different lengths, provided that the CFO estimation errors follow the wrapped normal distribution \cite{wang2015maximum}.
Through theoretical analysis, we show that the distribution of the CFO estimation error can be well approximated by the wrapped normal distribution and the CCMLE can achieve the Cram$\acute{\text{e}}$r-Rao Bound (CRB) when a certain signal-to-noise ratio (SNR) threshold is met. Meanwhile, this threshold is given in an approximate analytic form.
Moreover, we present a parameter configuration guideline based on these theoretical analyses to achieve a better MSE performance and a lower SNR threshold.
Furthermore, simulation experiments demonstrate high consistency with the theoretical analysis and show that our proposed CCMLE method has advantages in terms of estimated range and error performance compared to baseline schemes.
The main contributions of this article are presented in the following:
\begin{itemize}[itemsep=0pt,topsep=0pt,parsep=0pt]
    \item{We first show that the distribution of the CFO estimation error can be modeled by the von Mises distribution \cite{jangsher2022ber, jang2015bit}, which is a close approximation to the wrapped normal distribution \cite{mardia2000directional}.
        And then, we propose the CCMLE method to achieve the optimal performance for high-mobility OFDM systems, which has the advantages of a full estimated range and low additional complexity.}
    \item{We analyze the CRB of the CFO estimation under varying CFO estimated ranges, and reveal that the theoretical mean square error (MSE) of the proposed CCMLE method is actually the CRB.
    Besides, we give the analytic expression for the SNR threshold approaching the CRB, which enables an efficient waveform design.}
    \item{With the comprehensive analysis of the SNR threshold approaching the CRB and theoretical MSE, we give a guideline for the parameter configurations of the proposed CCMLE method to achieve a better MSE performance and a lower SNR threshold.}
\end{itemize}

The rest of the paper is organized as follows. Section \ref{System Model} presents the problem statement and the system model.
In Section \ref{S-CCMLE}, we formulate the CFO estimation problem as the remaindering problem and discuss the distribution of the CFO estimation errors. And then, the CCMLE method is proposed based on the MLE-based CRT.
Section \ref{Guide} shows the performance analysis of the proposed CCMLE method and gives a parameter configuration guideline for this method.
Furthermore, we present a brief complexity analysis for the proposed CCMLE and baseline methods in Section \ref{Complexity}.
Next, the experiment results are shown in Section \ref{EXPERIMENT}.
Finally, the conclusions are given in Section \ref{Conclusion}.

\section{Problem Statement and System Model}\label{System Model}
We consider a high-mobility scenario where the line-of-sight (LOS) path is dominant \cite{fontan2001statistical,loo1985statistical}.
As a result, the communication system will suffer from severe LOS component Doppler shifts due to high mobility.
For example, a high-speed aircraft using an OFDM system will suffer from severe CFOs, i.e., LOS component Doppler shifts, that significantly exceed the subcarrier spacing.
Meanwhile, we assume that the speed of the aircraft and the angle of the received signal are quasi-static for a certain period of time. For example, a fifth generation (5G) frame of 10ms in length \cite{shafi20175g}, during which the above quasi-static assumption is reasonable.
Moreover, we assume that the transmitted time-domain training symbols $s(n)$ consist of the ZC sequences and satisfy
\begin{equation}
\begin{array}{l}
\displaystyle {s^*}(n)s(n) = {\left| {s(n)} \right|^2} = 1,\,{n = 0,1,2, \ldots},
\end{array}\label{eq2n}
\end{equation}
and
\begin{equation}
\begin{array}{l}
\displaystyle s(n+L)=s(n),\,{n = 0,1,2, \ldots,L-1},
\end{array}\label{eq2nb}
\end{equation}
where $n$ represents the discrete sample point and $L$ represents the sample interval.
In this case, the received time-domain signals $r(n)$ can be expressed as
\begin{equation}
\begin{array}{*{20}{c}}
{{r(n)} = h(n)s(n)e^{j2\pi(\varepsilon_N\Delta f)nT_s} + \omega(n),}
\end{array}\label{eq1}
\end{equation}
where $h(n)$ represents the channel coefficient of the LOS component, $\omega(n)$ is the additive white Gaussian noise (AWGN), $\varepsilon_N \in [0,N)$ is the normalized CFO with respect to subcarrier spacing $\Delta f=(NT_s)^{-1}$, $N$ is the discrete Fourier transform size, and $T_s$ is the sampling period. 
Note that, we consider the channel is constant during the period of training symbols.
Furthermore, to facilitate mathematical analysis, we assume that $\left|h(n)\right|^2 = 1$ by considering the quasi-static characteristic without loss of generality.

In conventional training symbol aided CFO estimation, the normalized CFO can be estimated by the phase difference between identical sample intervals \cite{schmidl1997robust}, described as
\begin{equation}
\begin{array}{l}
\displaystyle
{{\hat \varepsilon }_N} = \frac{N}{{2\pi L}} \text {arg}({P_L}),
\end{array}\label{eq2}
\end{equation}
where $\hat\varepsilon_N$ is the estimated normalized CFO, the function $\text {arg}(\cdot)$ returns the phase angle in radians, and ${P_L}$ is a correlation function with the sample interval $L$, defined as
\begin{equation}
\begin{array}{l}
\displaystyle
{P_L} = \sum\limits_{m = 0}^{L - 1} {{r^*}(m)r(m + L)}.
\end{array}\label{eq3}
\end{equation}

As a result, the estimated range of the normalized CFO is $[-N/2L,N/2L]$.
Intuitively, a smaller sample interval $L$ is required to obtain a wider range of estimates, which leads to a lower accuracy \cite{schmidl1997robust,huang2021synchronization}.


As mentioned before, we apply CRT to perform CFO estimation in this paper.
It is important to note that for the purpose of describing the CRT, we denote the estimated range as $\left[0,\frac{N}{{L}}\right]$ and one can return $\hat\varepsilon_N$ to $\left[-\frac{N}{{2L}},\frac{N}{{2L}}\right]$ using the following formula:
\begin{equation}
\begin{array}{l}
\displaystyle
\hat\varepsilon_N = \left((\hat\varepsilon_N + \frac{N}{{2L}}) \enspace \bmod \enspace \frac{N}{{L}}\right)-\frac{N}{{2L}},
\end{array}\label{eq4}
\end{equation}
where the notation `$\bmod$' represents modular arithmetic.
\section{Proposed MLE-based CRT Method for CFO Estimation}\label{S-CCMLE}

In order to apply the CRT in the CFO estimation problem, $K$ different estimated ranges $[0, \Gamma_1),\ldots,[0, \Gamma_K)$ are required, while the greatest common divisor (gcd) of any two different estimated ranges needs to be satisfied as 1, i.e., $gcd(\Gamma_i,\Gamma_j)=1$ for $i \ne j$. 
Without loss of generality, assume that the above relatively co-prime numbers satisfy $\Gamma_1 < \cdots < \Gamma_K$.
In addition, to ensure that the above estimated ranges can be obtained, the estimated CFO requires to be normalized to $(\Gamma T_s)^{-1}$ instead of $(N T_s)^{-1}$, where $\Gamma=\Gamma_1\Gamma_2\cdots\Gamma_K$.
Accordingly, the sample intervals for the training symbols are set to be $L_1,\ldots,L_K$, where $L_i=\Gamma/\Gamma_i$ for $i=1,\ldots,K$.

Therefore, the normalized CFOs of different sample intervals can be represented as
\begin{equation}
\begin{array}{l}
\displaystyle
{{\hat \varepsilon }_i} = \frac{\Gamma}{{2\pi L_i}}\text {arg}({P_{L_i}}),\, i=1,\ldots,K,
\end{array}\label{eq5}
\end{equation}
where {$P_{L_i}$ is obtained by replacing $L$ in (\ref{eq3}) with $L_i$}, ${{\hat \varepsilon }_i}$ is the estimated CFO normalized to $(\Gamma T_s)^{-1}$, and the corresponding estimated range is $[0, \Gamma_i)$.

Furthermore, we can rewrite (\ref{eq5}) into the following form:
\begin{equation}
\begin{array}{l}
\hat\varepsilon_i=\varepsilon_i+\Delta\varepsilon_i,\, i=1,\ldots,K,
\end{array}\label{eq9}
\end{equation}
where $\Delta\varepsilon_i$ represents the normalized CFO estimation error caused by the noise.

Let $\varepsilon$ be the normalized CFO to be estimated, which is normalized to $(\Gamma T_s)^{-1}$.
If we ignore the effect of noise temporarily, $\varepsilon_i$ will be exactly the remainder of $\varepsilon$ modulo $\Gamma_i$.
This is because the estimated phase by the CFO estimation has cyclic characteristics and the estimated range with sample interval $L_i$ used is exactly $[0, \Gamma_i)$. Hence, we can obtain the following equations on the condition that there is no noise,
\begin{equation}
\begin{array}{l}
\varepsilon _i  \equiv {\varepsilon}\bmod {\Gamma _i},\, i=1,\ldots,K.
\end{array}\label{eq10}
\end{equation}
where $0 \le \varepsilon_i < \Gamma_i$, denoted by $\varepsilon _i = {\left\langle {\varepsilon} \right\rangle _{\Gamma_i}}$.

As a result, the CFO estimation problem can be formulated as the remaindering problem for a real number $\varepsilon$, i.e., recovering the real number $\varepsilon$ from its erroneous remainders $\hat\varepsilon_i$, where the remainder noises are denoted as $\Delta\varepsilon_i$.

Based on the property of the CRT \cite{wang2015maximum}, $\varepsilon$ can be uniquely reconstructed if and only if $0 \le \varepsilon < \Gamma$.
In such case, a wide estimated range $[0, \Gamma)$ can be obtained from several different small estimated ranges $[0, \Gamma_i)$.

\subsection{Estimated Range of CFOs}
Based on (\ref{eq5}), since the CRT-based CFO estimation requires $\Gamma$ rather than $N$ as a normalized factor, the estimated normalized CFO is ${{\hat \varepsilon }}$ times virtual subcarrier spacing, given by
\begin{equation}
\Delta f_V=(\Gamma T_s)^{-1},
\end{equation}
rather than the actual subcarrier spacing
\begin{equation}
\Delta f=(N T_s)^{-1}.
\end{equation}
Hence, the relationship between $\Delta f_V$ and $\Delta f$ can be represented as
\begin{equation}
\begin{array}{l}
\displaystyle
\frac{\Delta f_V}{ \Delta f} = \frac{N}{\Gamma}.
\end{array}\label{eq6}
\end{equation}

As a result, the CFO normalized to $(NT_s)^{-1}$, i.e., $\hat\varepsilon_N$, can be obtained by
\begin{equation}
\begin{array}{l}
\displaystyle
\hat\varepsilon_N = \frac{N}{\Gamma} {{\hat \varepsilon }}.
\end{array}\label{eq7}
\end{equation}

Since the estimated range of ${{\hat \varepsilon }}$ is $[0, \Gamma)$, we can obtain a wide estimated range of the CFO normalized to $(NT_s)^{-1}$ from (\ref{eq7}), i.e., $0 \le {{\hat \varepsilon_N }} < N$.
Accordingly, based on (\ref{eq4}), the range of normalized CFOs that can actually be estimated is $[-N/2,N/2]$.
\vspace{-6pt}
\subsection{Normalized CFO Error Distribution}
In this section, we will analyze the distribution of errors for the estimated normalized CFO.
According to (\ref{eq5}), we know that the estimated error of $\hat\varepsilon_i$ is proportional to the estimated phase error of $\text {arg}(P_{L_i})$.\
Therefore, the problem of the estimated normalized CFO error distribution can be transformed into the analysis of the estimated phase error distribution.

Let $\hat{\theta_i}=\text {arg}(P_{L_i})$ denote the estimated phase, and then the phase error is defined as
\begin{equation}
\begin{array}{l}
\epsilon_i=\hat{\theta_i}-\theta_i.
\end{array}\label{eq10n}
\end{equation}
where $\theta_i$ is the actual phase.

Based on (\ref{eq1}) and (\ref{eq3}), we can rewrite the correlation function $P_{L_i}$ into the following form:
\begin{equation}
\begin{array}{l}
{P_{L_i}} = {L_i}{e^{j2\pi L_i(\varepsilon_N\Delta f){T_s}}} + {\omega_{L_i}},
\end{array}\label{eq8}
\end{equation}
where $\omega_{L_i}$ is still the AWGN based on the Central Limit Theorem (CLT).

According to \cite{jangsher2022ber}, if $\omega_{L_i}$ is the AWGN, the probability distribution function (PDF) of the phase error $\epsilon_i$ can be perfectly matched by the von Mises distribution over a wide range of SNRs.
{Therefore, we have that the phase error $\epsilon_i$ follows the von Mises distribution  with mean zero.}

According to \cite{mardia2000directional}, the von Mises distribution is a close approximation to the wrapped normal distribution.
Furthermore, based on (\ref{eq5}), we know that the estimated normalized CFOs are linear transformations of the estimated phases.
{Therefore, it is reasonable to assume that the normalized CFO estimation error follows the wrapped normal distribution with mean zero.}

Additionally, even though the received signals have the same SNR, the normalized CFO estimates have different variances. This is because different sample intervals ($L_i$) are used in the correlation function. As a result, the normalized CFO estimation errors obtained from different sample intervals result in a wrapped normal distribution with varying variances.
\vspace{-6pt}
\subsection{CRT-based CFO Maximum Likelihood Estimation}\label{CCMLE}

Since the normalized CFO estimation errors, i.e., the remainder noises of robust CRT problem, follow the wrapped normal distribution with varying variances, we propose to use the MLE-based robust CRT \cite{wang2015maximum} to obtain optimal performance.
This proposed CRT-based CFO MLE (CCMLE) method is detailed in the following.

Let $M_i=M\Gamma_i$,where $M$ is an integer greater than or equal to 2. Substitute $\Gamma$ in (\ref{eq5}) by $M\Gamma$ and $\Gamma_i$ in (\ref{eq10}) by $M_i$, we have
\begin{equation}
\begin{array}{l}
\varepsilon_{M_i}  \equiv {\varepsilon _M}\bmod {M _i},\, i=1,\ldots,K,
\end{array}\label{eq14n}
\end{equation}
where $\varepsilon_M $ and ${\varepsilon _{M_i}}$ are the CFOs normalized to $(M\Gamma T_s)^{-1}$, and the corresponding range are $[0, M\Gamma)$ and $[0, M\Gamma_i)$.
Hence, similar to (\ref{eq9}), we have the following form:
\begin{equation}
\begin{array}{l}
\hat\varepsilon_{M_i}=\varepsilon_{M_i}+\Delta\varepsilon_{M_i},\, i=1,\ldots,K.
\end{array}\label{eq15n}
\end{equation}

As described in \cite{schmidl1997robust}, by using the method in \cite{moose1994technique}, the variance of the CFO estimates for different-length sample intervals is found to be
\begin{equation}
\begin{array}{l}
\displaystyle
\sigma^2_i = \frac{{{M^2\Gamma ^2}}}{{4{\pi ^2}L_i^3 \cdot \eta}},
\end{array}\label{eq11}
\end{equation}
where $\sigma^2_i$ ($i=1,\ldots, K$) represents the variance of the normalized CFO estimation error $\Delta\varepsilon_{M_i}$ and $\eta$ represents the SNR of the received signals.

Therefore, we can conclude that the remainder noise $\Delta\varepsilon_{M_i}$ follows the wrapped normal distribution with mean zero and variance $\sigma^2_i$ for $i=1,\ldots, K$.

To optimally reconstruct $\varepsilon_M $, it is essential to optimally determine the common remainder $r^c$ from erroneous common remainders $\hat r_i^c$.
First, $\hat r_i^c$ is obtained by noisy remainder $\hat\varepsilon_{M_i}$ modulo $M$, i.e.,
\begin{equation}
\begin{array}{l}
\displaystyle \hat r_i^c = {\left\langle {{{\hat \varepsilon }_{{M_i}}}} \right\rangle _M}, \, i=1,\ldots,K.
\end{array}\label{eq18n}
\end{equation}

Next, the optimal estimate for the common remainder $r^c$ can be determined by the following equation \cite{wang2015maximum}:
\begin{equation}
\begin{array}{l}
\displaystyle \hat r^c = \arg \mathop {\min }\limits_{x \in \Omega } \sum\limits_{i = 1}^K {{w_i}d_M^2\left( {\hat r_i^c,x} \right)},
\end{array}\label{eq19n}
\end{equation}
where $w_i$ is the weights of the remainders defined as
\begin{equation}
\begin{array}{l}
\displaystyle {w_i} = \frac{{\frac{1}{\sigma _i^2}}}{{\sum\limits_{i = 1}^K \frac{1}{\sigma _i^2} }}, \, i=1,\ldots,K.
\end{array}\label{eq15}
\end{equation}


\noindent $d_M\left( {\hat r_i^c,x} \right)$ is the circular distance defined as
\begin{equation}
\begin{array}{l}
\displaystyle d_M\left( {\hat r_i^c,x} \right) \buildrel \Delta \over = \hat r_i^c-x-\left[ \frac{\hat r_i^c-x}{M} \right]M,
\end{array}\label{eq21n}
\end{equation}
where $[\cdot]$ stands for the rounding integer.
Besides, the set of the optimal solutions $\Omega$ is defined as
\begin{equation}
\Omega = \left\{ {{{\left\langle {\sum\limits_{i = 1}^K {{w_i}\hat r_i^c + M\sum\limits_{i = 1}^t {{w_{\rho (i)}}} } } \right\rangle }_M},t = 1, \ldots ,K} \right\},
\label{eq23n}
\end{equation}
where $\rho$ is a permutation of the set $\left\{1,\ldots,K \right\}$ such that
\begin{equation}
\begin{array}{l}
\displaystyle \hat r_{\rho _{(1)}}^c \le  \cdots  \le \hat r_{\rho _{(K)}}^c,
\end{array}\label{eq24n}
\end{equation}
and $w_{\rho _{(i)}}$ is the weight of $\hat r_{\rho _{(i)}}^c$.

\begin{algorithm}[!t]
\caption{The proposed CCMLE method}
\label{alg1}
\hspace*{0.02in} {\bf Input:} $M,\Gamma_i,r(n)$ \\
\hspace*{0.02in} {\bf Output:}  $\hat \varepsilon_N$
\begin{algorithmic}[1]
\State Initialize $\Gamma=\Gamma_1\Gamma_2\cdots\Gamma_K$, $L_i=\Gamma/\Gamma_i$, and initialize $\bar L_i$ according to (\ref{eq26n}).
\State Calculate ${\hat \varepsilon }_{{M_i}},\sigma_i^2$ ($i=1,\ldots, K$) according to (\ref{eq5}) and (\ref{eq11}), respectively.
\State Calculate $\hat r_i^c$ ($i=1,\ldots, K$) according to (\ref{eq18n}).
\State Construct the set of optimal solutions $\Omega$ according to (\ref{eq23n}).
\State Determine the optimal common remainder $\hat r^c$ from $\Omega$ according to (\ref{eq19n}).
\State Calculate ${\hat \varepsilon }_{{M}}$ according to (\ref{eq25n}).
\State Convert ${\hat \varepsilon }_{{M}}$ to ${\hat \varepsilon }_{{N}}$ according to (\ref{eq7n}).
\State \Return ${\hat \varepsilon }_{{N}}$.
\end{algorithmic}
\end{algorithm}

After the common remainder $r^c$ is optimally determined, the normalized CFOs can be estimated by
\begin{equation}
\begin{array}{l}
\hat \varepsilon_M = M\left( \left(\sum\limits_{i = 1}^K {{{\bar L}_i}{L_i}{{\hat q}_i}}\right) \enspace \bmod \enspace \Gamma \right) + \hat r^c,
\end{array}\label{eq25n}
\end{equation}
where
\begin{equation}
{\hat q}_i = \left[ \frac{{\hat \varepsilon }_{{M_i}}-\hat r_i^c}{M} \right],\, i=1,\ldots,K.
\label{eq27n}
\end{equation}
$\bar L_i$ are predetermined constants, i.e., the modular multiplicative inverse of $L_i$ modulo $\Gamma_i$, defined as
\begin{equation}
{\bar L}_i{L_i}\equiv 1 \bmod {\Gamma _i},\, i=1,\ldots,K.
\label{eq26n}
\end{equation}

Finally, based on (\ref{eq7}), the CFO normalized to $(NT_s)^{-1}$, i.e., $\hat\varepsilon_N$, is given by
\begin{equation}
\begin{array}{l}
\displaystyle
\hat\varepsilon_N = \frac{N}{M\Gamma} {{\hat \varepsilon_M }}.
\end{array}\label{eq7n}
\end{equation}

\begin{figure}[!t]
\centerline{\includegraphics[width=2.8 in]{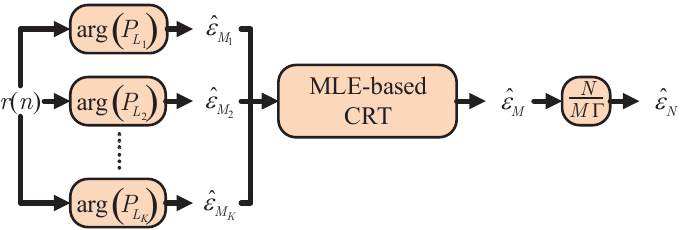}}
\caption{Diagram for the proposed CCMLE method.}
\label{Diagram}
\end{figure}

We summarize the proposed CCMLE method as Algorithm \ref{alg1} and the corresponding diagram is given in Fig. \ref{Diagram}.

\section{Performance Analysis and Parameter Configuration Guidelines}\label{Guide}
\subsection{CFO Estimation Performance Analysis}
According to \cite{wang2015maximum}, the theoretical MSE of $\hat\varepsilon_{M}$ for the proposed CCMLE method is given by
\begin{equation}
\begin{array}{l}
\displaystyle \Delta_{MSE(M)} = \sum\limits_{i = 1}^K {w_i^2} \sigma _i^2.
\end{array}\label{eq16}
\end{equation}

For the CFO estimation performance, we have the following theorem.

\textit{Theorem 1:} When the weights $w_i$ are defined by (\ref{eq15}), the $\Delta_{MSE(M)}$ in (\ref{eq16}) is smaller than the minimum of $\sigma^2_i$ ($i=1,\ldots, K$) defined by (\ref{eq11}), i.e., $\Delta_{MSE(M)} < \rm{min}\{\sigma^2_{\it{i}}\}$, where $\rm{min}\{\sigma^2_{\it{i}}\}$ represents the minimum of the set $\{\sigma^2_{\it{i}},i=1,\ldots, K\}$.
Furthermore, the CRB $var[\hat\varepsilon_{M}]$ of the normalized CFO estimates $\hat\varepsilon_{M}$ with $K$ different sample intervals $L_1,\ldots,L_K$ used is given by
\begin{equation}
\begin{array}{l}
\displaystyle
var[\hat\varepsilon_{M}] \ge \Delta_{MSE(M)}.
\end{array}\label{eq23}
\end{equation}
\textit{Proof:} The proof of this theorem is relegated to Appendix \ref{appendixT1}.

{This theorem states that the performance of the proposed CCMLE method is better than that obtained under any of the single-length sample intervals used.
Meanwhile, the normalized CFO estimated errors obtained by the proposed CCMLE method can approach the CRB.}

Intuitively, based on (\ref{eq7n}) and (\ref{eq16}), the theoretical MSE of $\hat\varepsilon_{N}$ is given by
\begin{equation*}
\displaystyle \Delta_{MSE} = \left(\frac{N}{M\Gamma}\right)^2 \Delta_{MSE(M)}.
\end{equation*}
That is,
\begin{equation}
\begin{aligned}
\displaystyle \Delta_{MSE}=\left(\frac{N}{M\Gamma}\right)^2 \sum\limits_{i = 1}^K {w_i^2} \sigma _i^2.
\end{aligned}\label{eq16n}
\end{equation}
Accordingly, the CRB of $\hat\varepsilon_{N}$ is given by
\begin{equation}
\begin{array}{l}
\displaystyle
var[\hat\varepsilon_{N}] \ge \Delta_{MSE}.
\end{array}\label{eq40n}
\end{equation}

{It can be seen from (\ref{eq23}) and (\ref{eq40n}) that the MSE of the proposed CCMLE method can approach the CRB, which indicates that the proposed CFO estimator can achieve the optimal performance.
{In addition, we show in the following theorem that given the values of $\Gamma_1,\Gamma_2,\ldots,\Gamma_{K}$, the proposed CCMLE is theoretically capable of yielding a SNR threshold approaching the CRB.}
Before that, we give the following \textit{Lemma 1} to facilitate the derivation of \textit{Theorem 2}.}

\textit{Lemma 1:} If $L_1 > \cdots > L_K>0$, and $\xi_\Psi$ is defined as

\begin{equation}\label{eq48}
{\xi _\Psi } = \left\{ {\begin{array}{*{20}{l}}
\displaystyle{\frac{1}{{\sum\limits_{\Delta {\varepsilon _{{M_i}}} \in S} \!{L_i^3} }} + \frac{1}{{\sum\limits_{\Delta {\varepsilon _{{M_j}}} \in \bar S} \!{L_j^3} }},}&{\!\!\!\!\!\!S \ne \emptyset ,\bar S \ne \emptyset }\vspace{1ex}\\
\displaystyle {\frac{1}{{\sum\limits_{i = 1}^K {L_i^3} }},}&{\!\!\!\!\!\!S = \emptyset {\rm{\, or \,}}\bar S = \emptyset }
\end{array}} \right.\!\!\!\!,
\end{equation}
where $S$ is the subset of $U=\left\{{{\Delta \varepsilon }_{{M_1}}}, \ldots,{{\Delta \varepsilon }_{{M_K}}}\right\}$, and $\bar S$ is the complement of $S$ in $U$.
Then, the maximum value of $\xi_\Psi$ is given by
\begin{equation*}\label{eq53bb}
\xi_\Psi^*={\frac{1}{L_K^3}} + {\frac{1}{{\sum\limits_{j\ne K} {L_j^3} }}}.
\end{equation*}
\textit{Proof:} The proof of this theorem is relegated to Appendix \ref{appendixA}.


\textit{Theorem 2:} For an arbitrarily small real number $\delta$, MSE of the CCMLE method approaches the CRB with a probability of at least $1-\delta$ for 
\begin{equation}\label{eq53aa}
\eta\ge\frac{\Gamma^2 x_\delta^2 \xi_\Psi^*}{\pi^2},
\end{equation}
{where $x_\delta$ is a constant with respect to $\delta$, and $\xi_\Psi^*$ is given by (\ref{A12}).}\\
\textit{Proof:} The proof of this theorem is relegated to Appendix \ref{appendixT2}.

\subsection{Parameter Configuration Guideline}
Next, based on \textit{Theorem 1} and \textit{Theorem 2}, we thoroughly examine how the parameter configurations affect the performance of the CCMLE method, taking into account the MSE and the SNR threshold $\eta_{th}$.
After that, we present a guideline for the parameter configurations of the proposed CCMLE method to achieve a better MSE performance and a lower SNR threshold.

First, based on (\ref{eq50}) and (\ref{eq53}), we know that $\eta_{th}$ is determined by the value of $\Gamma^2 \xi_\Psi^*$ for a given arbitrarily small number $\delta$.
Recall that $\Gamma=\Gamma_1\Gamma_2\cdots\Gamma_K$ and $L_i=\Gamma/\Gamma_i$.
Then, based on the fact that $\Gamma_1 < \cdots < \Gamma_K$ and $\Gamma_1,\Gamma_2,\ldots,\Gamma_{K}$ are co-prime numbers, we have $L_K^3 \ll {\sum\limits_{j\ne K} {L_j^3} }$.
As a result, based on (\ref{A12}), the value of $\Gamma^2 \xi_\Psi^*$ can be approximated as
\begin{equation}\label{eq48b}
\begin{aligned}
\Gamma^2 \xi_\Psi^*\approx{\frac{\Gamma_K^2}{\Gamma_1\Gamma_2\cdots\Gamma_{K-1}}}.
\end{aligned}
\end{equation}

It can be seen from (\ref{eq48b}) that the larger the $\Gamma_1\Gamma_2\cdots\Gamma_{K-1}$, the smaller the $\Gamma^2 \xi_\Psi^*$ for a given $\Gamma_K$.
Furthermore, because $\Gamma_1,\Gamma_2,\ldots,\Gamma_{K}$ are co-prime numbers, the maximum value of $\Gamma_1\Gamma_2\cdots\Gamma_{K-1}$ is obtained when $\Gamma_1,\Gamma_2,\ldots,\Gamma_{K-1}$ are consecutive $K-1$ primes less than $\Gamma_K$.

Moreover, according to (\ref{eq30}) and (\ref{eq16n}), we can rewrite the theoretical MSE of $\hat\varepsilon_{N}$ as
\begin{equation}
\begin{aligned}
\displaystyle \Delta_{MSE} &= \frac{1}{\sum\limits_{i = 1}^K \frac{{{4{\pi ^2}L_i^3 \cdot \eta}}}{{N^2}}},
\end{aligned}\label{eq30a}
\end{equation}
which shows that the larger $K$ and $L_i$ will result in a smaller $\Delta_{MSE}$ for a given SNR.
Besides, based on (\ref{eq48b}) and (\ref{eq30a}), it can be found that the value of $M$ has no effect on the final performance.
Thus, without loss of generality, we set $M=2$ in the following simulation.

In general, a group of parameter configurations will be selected to achieve a better MSE performance and a lower SNR threshold, leading to the following configuration guidelines.

\textit{Guidelines: } We should use as many $K$ different estimated ranges as possible, and set $\Gamma_i(i=1,\ldots,K)$ to be $K$ consecutive co-prime numbers, while making the sample intervals $L_i(i=1,\ldots,K)$ as large as possible.

\section{Complexity Analysis}\label{Complexity}
In this section, we compare the complexity of the proposed CCMLE and baseline methods.
The main benchmarks for comparison can be divided into two categories: CRT-based methods and traditional methods.
Specifically, the CRT-based methods include the closed-form robust CRT-based method \cite{huang2018doppler,wang2010closed}, denoted as ``Closed-form CRT'', and the classic CRT-based method \cite{rha2015real}, abbreviated as ``Classic CRT''. 
For traditional methods, the typical two-stage estimation method that the IFO estimation based on the symmetrical correlation property of FFO-compensated received signal \cite{huang2021synchronization} is adopted for comparison, marked as ``Sym.-corr.''.
Besides, the Moose's method \cite{moose1994technique} is used for comparison as well.

All these methods obtain an initial CFO estimate based on the training symbols by means of autocorrelation and summation, and thus the corresponding complexity is equivalent to $O(L)$, where $L$ is the length of the sample interval.
Additional computation for the Classic CRT requires $(K-1)$ multiplications and $(K-1)$ additions.
For the two-stage estimation methods in \cite{huang2021synchronization}, additional computation is mainly involved in performing $N$ times autocorrelation and summation of the sequence and finding the minimum values, so the additional complexity is equivalent to $O(NL)$.
Meanwhile, our proposed CCMLE method needs additional computations to obtain  $\hat r_i^c$, $\hat r^c$, $\Omega$, $\hat q_i$ and $\hat \varepsilon_M$, but the overall complexity is equivalent to $O(K^2)$, which is essentially determined by the number of $K$.
Recall that $K$ represents the number of different estimated ranges, which is generally a small number, typically 3 or 4.
Therefore, only a small number of additional computations are required for the proposed CCMLE method.
However, since the Closed-form CRT estimates $r^c$ by searching all the reals in the range of $[0,M)$ \cite{wang2015maximum}, the complexity is inversely proportional to the step size $\lambda$ in search.
The complexity for the above methods is summarized in TABLE \ref{table2}.

\vspace{-6pt}
\begin{table}[!htp]
\caption{Complexity comparison}
\vspace{-6pt}
\tabcolsep=0.2cm
\renewcommand\arraystretch{1.6}
\begin{center}
\begin{tabular}{c|c}
  \hline
      & Complexity  \\ \hline
  CCMLE & $O(L)$ + $O(K^2)$\\ \hline
  Classic CRT \cite{rha2015real} & $O(L)$ + $O(K)$\\ \hline
  Closed-form CRT\cite{huang2018doppler} & $O(L)$ + $O(MK^2 / \lambda)$\\ \hline
  Sym.-corr.\cite{huang2021synchronization} & $O(L)$ + $O(NL)$\\ \hline
  Moose\cite{moose1994technique} & $O(L)$\\ \hline
\end{tabular}
\end{center}\label{table2}
\end{table}

\section{Experimental Results}\label{EXPERIMENT}
%
In this section, we demonstrate the performances of the proposed CCMLE method and the benchmarks in a high-mobility OFDM system.
Specifically, we consider the scenario that the maximum normalized CFO is $N/2$, i.e., $\varepsilon_N \in [-N/2,N/2]$, where $N$ is set to be 64 \footnote{
It is well-known that LEO satellite communication systems, such as Telesat, OneWeb, and SpaceX, use the 17.8-19.3 GHz bands for downlink communications \cite{del2019technical} and that LEO satellites move at very high speeds, typically 7.5 km/s \cite{phipps2014laser,chen2022coordinative}. Hence, the maximum Doppler shift can be as high as 480 kHz under the condition that the carrier frequency of the simulation is set to 19.2 GHz.
Furthermore, a commonly used subcarrier spacing of 15kHz is assumed \cite{huang2021synchronization}, and then the maximum normalized Doppler shift can be as high as 32 subcarriers.}
For parameter configurations of the proposed CCMLE method, we restrict the maximum sample interval $L_1$ to be less than $N$.
As a result, according to the configuration guidelines in Section-\ref{Guide}, we set $\Gamma_1=3$, $\Gamma_2=5$, $\Gamma_3=7$, corresponding to $L_1=35$, $L_2=21$, $L_3=15$.
It is worth noting that all the baselines are restricted to use no more than the maximum sample interval of the CCMLE method, i.e., $L_1$, for fairness of comparison.
Furthermore, we carry out $1 \times 10^6$ trials for each method.

\subsection{MSE Performance}

\begin{figure*}[!t]
	\centering
	\begin{subfigure}{0.355\linewidth}
		\centering
		\includegraphics[width=0.9\linewidth]{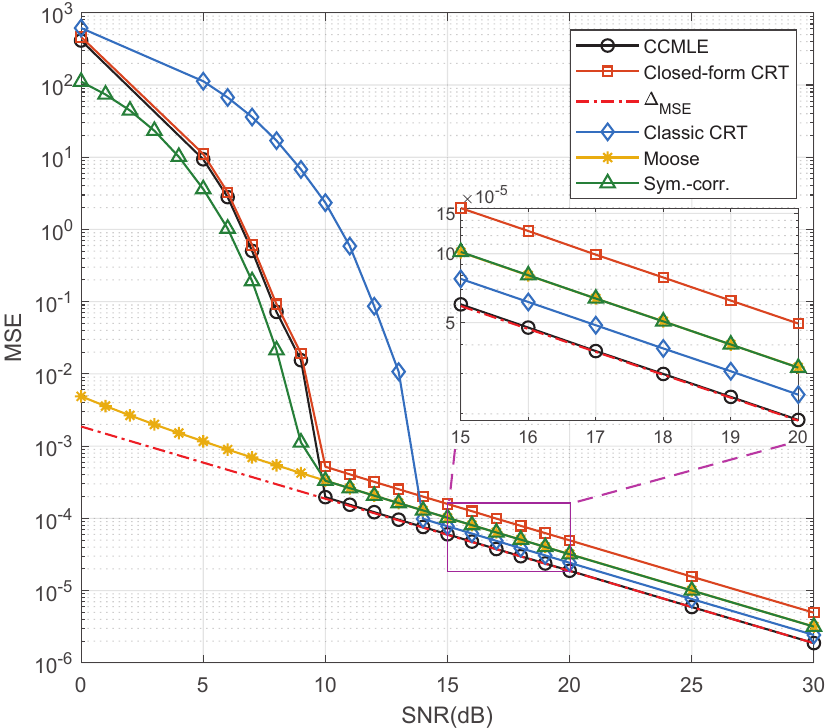}
		\caption{$\varepsilon_N=0.1$}
		\label{sim64a}
	\end{subfigure}
    \hspace{-9mm}
	\centering
	\begin{subfigure}{0.355\linewidth}
		\centering
		\includegraphics[width=0.9\linewidth]{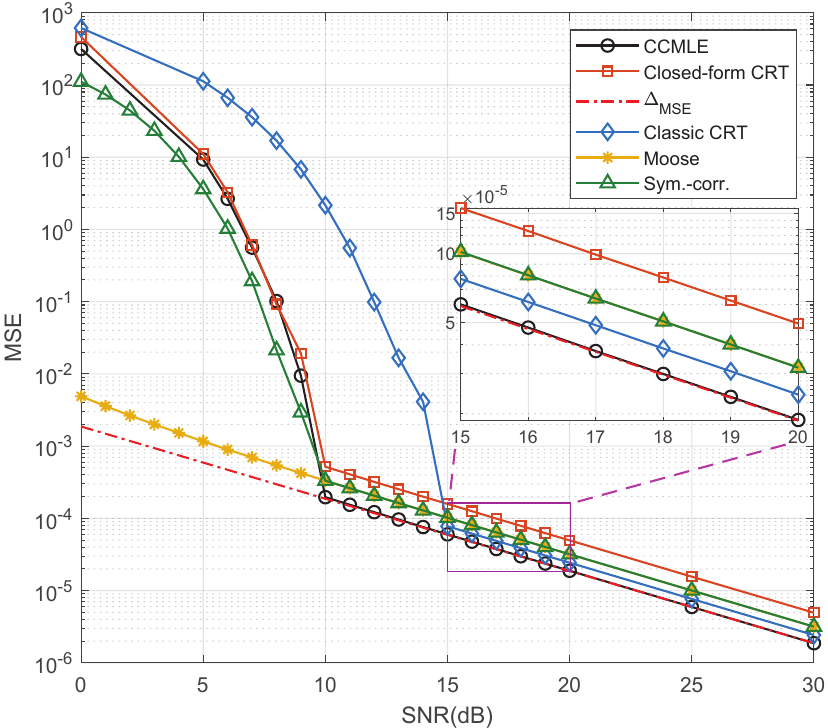}
		\caption{$\varepsilon_N=0.6$}
		\label{sim64b}
	\end{subfigure}
    \hspace{-9mm}
	\centering
	\begin{subfigure}{0.355\linewidth}
		\centering
		\includegraphics[width=0.9\linewidth]{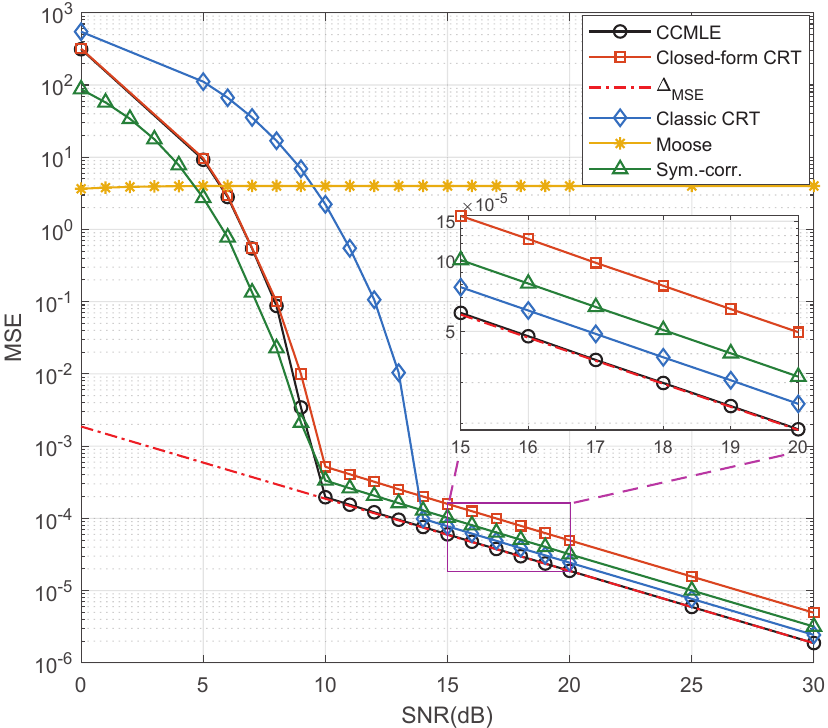}
		\caption{$\varepsilon_N=1.1$}
		\label{sim64c}
	\end{subfigure}

	\centering
	\begin{subfigure}{0.355\linewidth}
		\centering
		\includegraphics[width=0.9\linewidth]{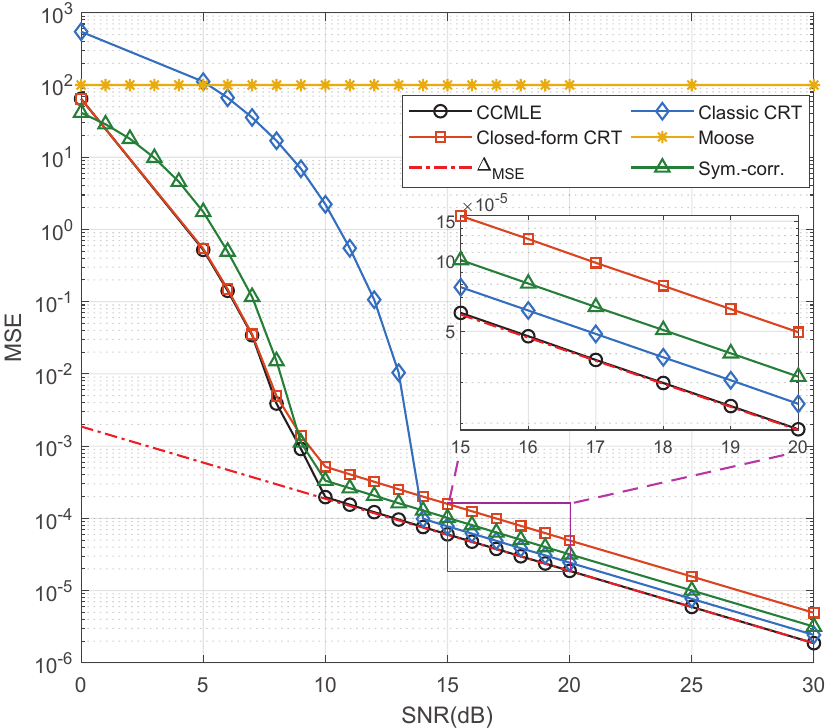}
		\caption{$\varepsilon_N=10.1$}
		\label{sim64d}
	\end{subfigure}
    \hspace{-9mm}
	\centering
	\begin{subfigure}{0.355\linewidth}
		\centering
		\includegraphics[width=0.9\linewidth]{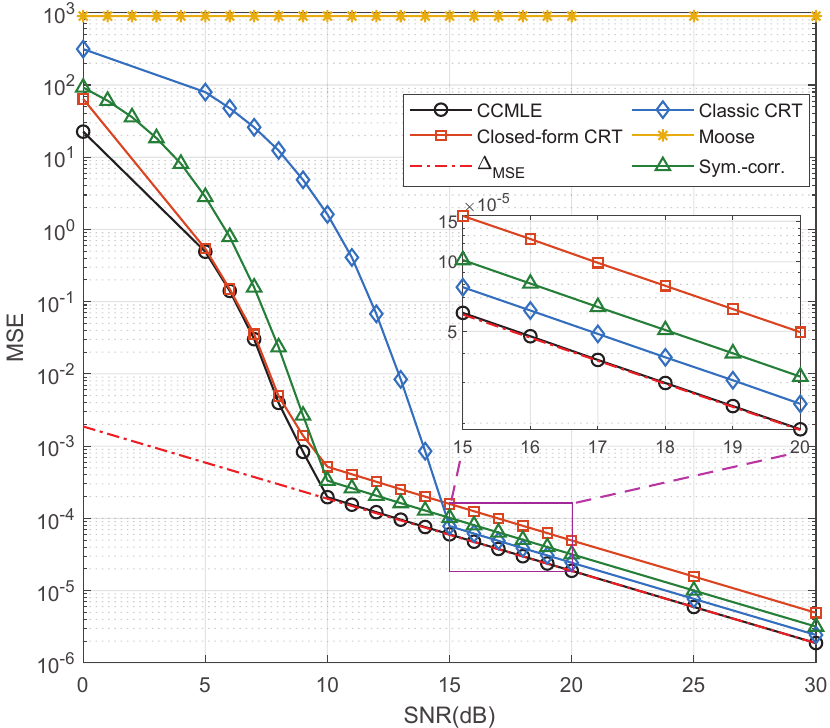}
		\caption{$\varepsilon_N=30.1$}
		\label{sim64e}
	\end{subfigure}
    \hspace{-9mm}
	\centering
	\begin{subfigure}{0.355\linewidth}
		\centering
		\includegraphics[width=0.9\linewidth]{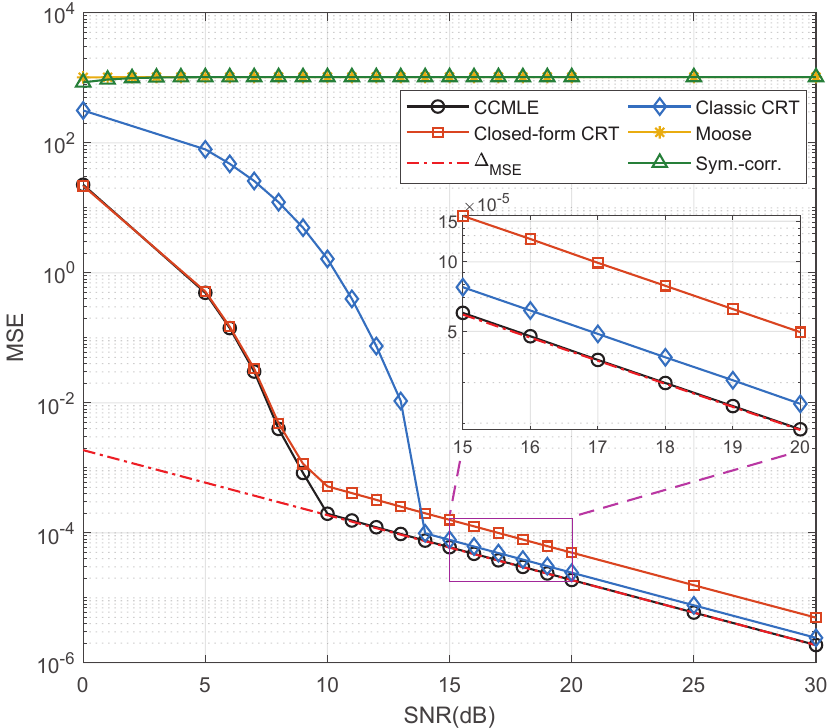}
		\caption{$\varepsilon_N=31.1$}
		\label{sim64f}
	\end{subfigure}
	\caption{Performances of the proposed method and baselines with fixed normalized CFOs ($N=64$)}
	\label{sim64}
\end{figure*}

We present the MSE performances of various fixed normalized CFOs, including $\varepsilon_N=$ 0.1, 0.6, 1.1, 10.1, 30.1, 31.1, for the proposed and baseline methods in Fig. \ref{sim64} from (a) to (f).
Note that, according to \textit{Theorem 1}, the CRB of the proposed CCMLE method corresponds to the theoretical MSE, i.e., $\Delta_{MSE}$.
The simulation results from Fig. \ref{sim64} (a) to (f) show that when the SNR is no smaller than 10dB, the proposed CCMLE approaches the CRB.
Besides, it can be found that the CCMLE achieves the best performance when the normalized CFO $\varepsilon_N$ increases from 0.1 to 31.1, in terms of a lower SNR threshold approaching the CRB and a smaller MSE.

{In addition, it can be seen from Fig. \ref{sim64} that the Closed-form CRT can reach the achievable performance of the algorithm at the same low SNR as the CCMLE, but its performance is still worse than the CCMLE.
This is due to the fact that both the CCMLE and the Closed-form CRT are derived from the robust CRT problem, but the latter does not take into account the cases that the remainder noise variances vary in different sample intervals, which results in a performance degradation that deviates from the CRB.
Moreover, Fig. \ref{sim64} shows that the achievable performance of the Classic CRT is always worse than $\Delta_{MSE}$.
This can be explained as follows.
According to \cite{rha2015real}, we know that the performance of the Classic CRT is upper bounded by the maximum sample interval $L_1$.
Therefore, based on \textit{Theorem 1}, we have that the achievable performance of the Classic CRT is always worse than the CRB of the CCMLE, i.e., $\Delta_{MSE}$.
Furthermore, it also can be seen from Fig. \ref{sim64} that the Classic CRT requires higher SNR to reach the achievable performance compared with the CCMLE.}

{For traditional methods, Fig. \ref{sim64} shows that Moose's method can perform much better than all the other methods in the low SNR range, i.e., $\rm{SNR}< 10\,\rm{dB}$, but its estimated performance degrades rapidly in all these SNR ranges at $\varepsilon_N > 1$.
This is because Moose's method only has an estimated range of $[-1,+1]$ \cite{schmidl1997robust}, indicating that it can work only in a small range of $\varepsilon_N$.
In addition, it can be seen from Fig. \ref{sim64} that the Sym.-corr. can reach the achievable performance at a similar SNR threshold as the CCMLE, but the achievable performance of the Sym.-corr. is still worse than $\Delta_{MSE}$.
Meanwhile, it can be found that the MSE of Sym.-corr. has a similar performance for most cases except when $\varepsilon_N = 31.1$.
It is worth noting that, Fig. \ref{sim64}(f) shows that the MSE performance of Sym.-corr. is similar to Moose's method for $\varepsilon_N = 31.1$, i.e., estimates are almost all failures, which indicates that Sym.-corr. could not work effectively for extremely large $\varepsilon_N$.}

\begin{figure}[!t]
\centerline{\includegraphics[width=3.5 in]{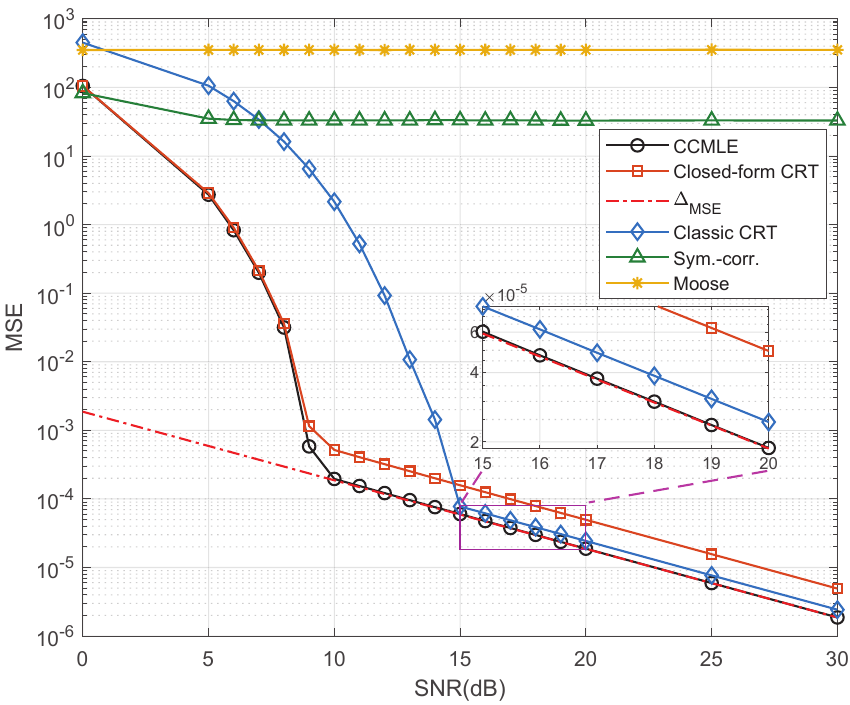}}
\caption{Performance comparison with $\varepsilon_N \in [-N/2,N/2]$ for different methods ($N=64$).}
\label{sim64all}
\end{figure}

Furthermore, to compare the performance over a wide range of the normalized CFO, we show the MSE performance for $\varepsilon_N $ sampled uniformly from $ [-N/2,N/2]$ in Fig. \ref{sim64all}.
The simulation results show a similar performance trend to that with fixed normalized CFOs.
However, it is worth noting that Moose's method and Sym.-corr. perform poorly over the entire simulated SNR range.
This is caused by the fact that the method in \cite{huang2021synchronization} is based on the correlation property of two symmetric sequences for IFO estimation.
However, this sequence will have the same properties as the original sequence if it is cyclically shifted by $N/2$.
Thus, this method is extremely prone to estimation ambiguity when the normalized CFO is close to $N/2$ or $-N/2$, i.e., boundaries of the estimated range, which will result in poor MSE performance.

\begin{figure}[!t]
\centerline{\includegraphics[height=3 in]{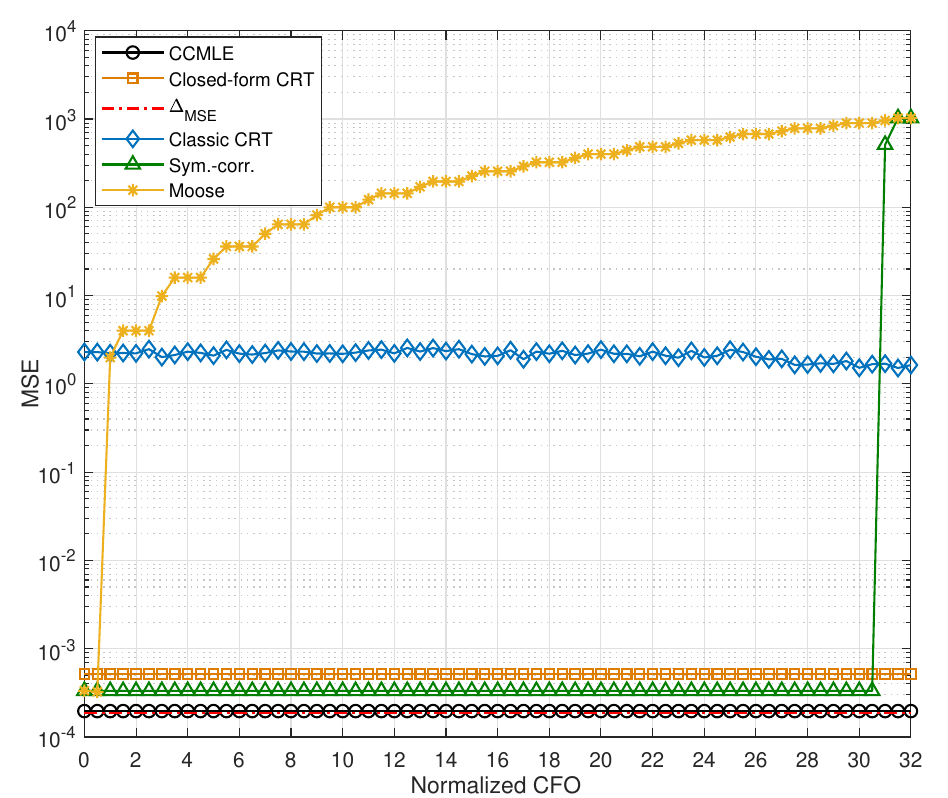}}
\caption{Performance comparison with an identical SNR for different methods ($N=64$, $\rm{SNR}= 10\,\rm{dB}$).}
\label{SNR10dB}
\end{figure}

{Further, we compare the MSE performance of these methods across different normalized CFO ranges and present their performances in the normalized CFO range of $ [0,N/2]$ in Fig. \ref{SNR10dB}.
We set $\rm{SNR}= 10\,\rm{dB}$, which is a SNR at which most methods reach their achievable performance, and the results in Fig. \ref{sim64} reflect this.
It can be seen from Fig. \ref{SNR10dB} that the proposed CCMLE can attain the CRB with $\varepsilon_N$ increasing from 0 to 32, which indicates that it has a full estimated range.
Furthermore, Fig. \ref{SNR10dB} shows that the Classic CRT and the Closed-form CRT also have the full estimated range, but they perform worse than the proposed CCMLE.
For traditional methods, it can be seen that Moose's method could not work effectively for $\varepsilon_N \ge 1$ and Sym.-corr. could not work effectively for $\varepsilon_N \ge 31$, which is consistent with the results in Fig. \ref{sim64}.
As a result, the CCMLE has the advantage of wider estimated range and better MSE performance compared with the benchmarks.}

\begin{figure}[!t]
\centerline{\includegraphics[height=3 in]{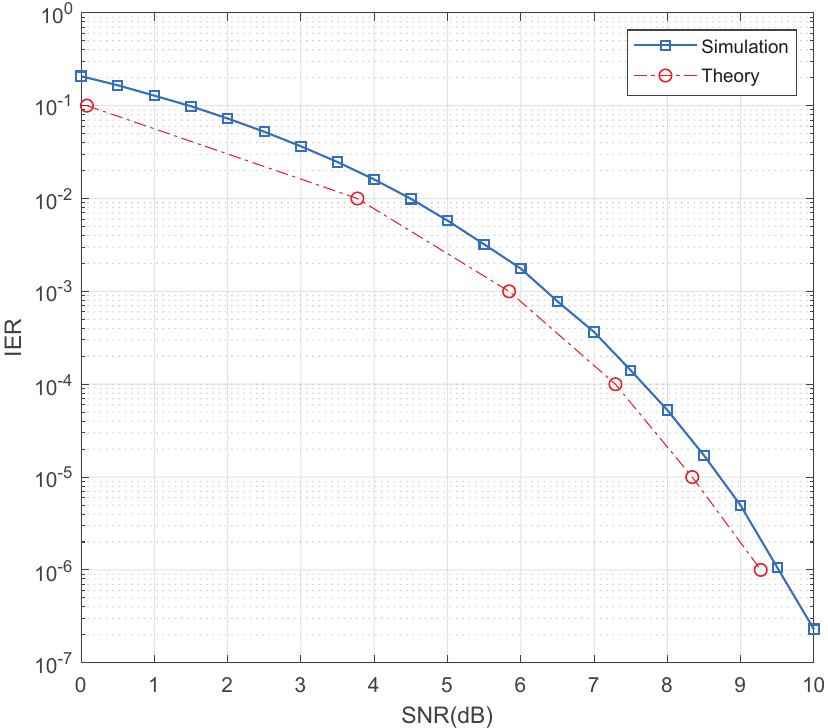}}
\caption{IER curve versus SNR for the proposed CCMLE method ($N=64$).}
\label{simTH}
\end{figure}

%

\subsection{SNR Threshold Evaluation}
In this section, we first theoretically give the SNR threshold approaching the CRB according to \textit{Theorem 2}. After that, the consistency of the theoretical analysis with the simulation results is presented.
In addition, an efficient waveform design is expected to be achieved through theoretical analysis.

First, due to the fact that $1 \times 10^6$ trials are carried out for different methods, it is reasonable to consider an arbitrarily small number $\delta=1\times 10^{-6}$ for \textit{Theorem 2}.
Hence, we can obtain $x_\delta \approx 4.9$ from TABLE \ref{table0} given $\delta=1\times 10^{-6}$.
Furthermore, according to (\ref{eq53}) and corresponding parameter configurations, we can obtain $\eta_{th} = 9.3 \,\rm{dB}$.
At the same time, it can be found from Fig. \ref{sim64} and Fig. \ref{sim64all} that the SNR thresholds approaching the CRB for the proposed CCMLE fall into the range $(9,\,10) \,\rm{dB}$,
which is consistent with the $9.3 \,\rm{dB}$ derived from \textit{Theorem 2}.


\begin{table}[!htbp]
\caption{Comparison between $\eta_{th}$ and $\eta_{\rm{sim}}$}
\renewcommand\arraystretch{1.6}
\begin{center}
\begin{tabular}{c|c|c|c}
  \hline
  IER               & $\eta_{th}$ (dB) & $\eta_{\rm{sim}}$ (dB) & $\eta_{\rm{sim}}-\eta_{th}$ (dB)\\ \hline
  $1\times 10^{-1}$ & 0   & 1.5 & 1.5 \\ \hline
  $1\times 10^{-2}$ & 3.8 & 4.5 & 0.7 \\ \hline
  $1\times 10^{-3}$ & 5.8 & 6.3 & 0.5 \\ \hline
  $1\times 10^{-4}$ & 7.3 & 7.7 & 0.4 \\ \hline
  $1\times 10^{-5}$ & 8.3 & 8.7 & 0.4 \\ \hline
  $1\times 10^{-6}$ & 9.3 & 9.5 & 0.2 \\
  \hline
\end{tabular}
\end{center}\label{tableTH}
\end{table}

Moreover, it can be found that the MSE values for low SNRs are always especially large, which is essentially caused by the estimation error of the integer part.
For example, in Fig. \ref{sim64all}, the MSE of the CCMLE is almost far greater than 1 when the $\rm{SNR}$ is less than $6 \,\rm{dB}$ and even reaches an order of magnitude of $10^2$ when the $\rm{SNR}$ is close to $0 \,\rm{dB}$.
Note that the MSE will deviates by orders of magnitude from theoretical value only when there is an IFO error.
In addition, the number of IFO errors decreases with increasing SNR until the CRB can be approached without IFO errors.
As a result, we can use the IFO error rate (IER) to reflect the probability of approaching the CRB for a given SNR.
In the simulation, IER is computed as follows,
\begin{equation}
\begin{aligned}
\displaystyle \text{IER}=N_{IE}/N_{all},
\end{aligned}\label{eqIER}
\end{equation}
where $N_{all}$ represents the total number of trials and $N_{IE}$ denotes the occurrence number of $|\hat \varepsilon_N-\varepsilon_N|>1$.
Note that the IER can be regarded as the arbitrarily small number $\delta$ in \textit{Theorem 2}, based on which the corresponding SNR threshold $\eta_{th}$ can be obtained.

\begin{figure*}[!t]
	\centering
	\begin{subfigure}{0.355\linewidth}
		\centering
		\includegraphics[width=0.9\linewidth]{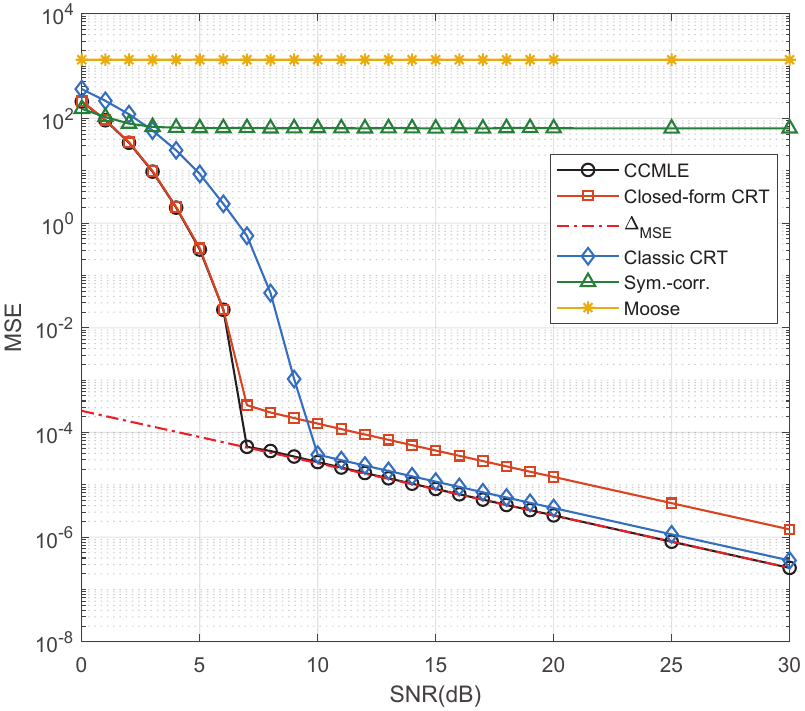}
		\caption{$N=128$}
		\label{N1}
	\end{subfigure}
    \hspace{-9mm}
	\centering
	\begin{subfigure}{0.355\linewidth}
		\centering
		\includegraphics[width=0.9\linewidth]{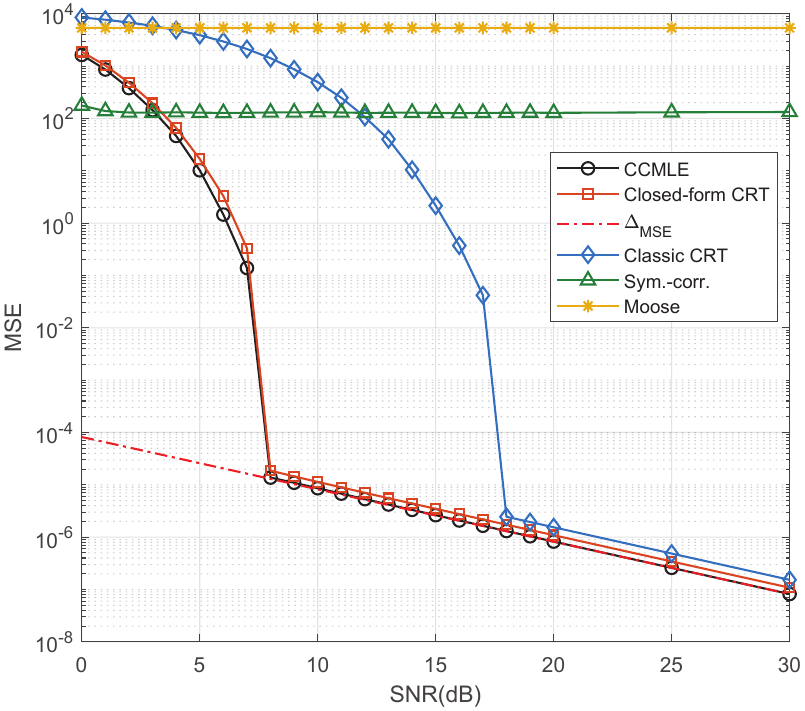}
		\caption{$N=256$}
		\label{N2}
	\end{subfigure}
    \hspace{-9mm}
	\centering
	\begin{subfigure}{0.355\linewidth}
		\centering
		\includegraphics[width=0.9\linewidth]{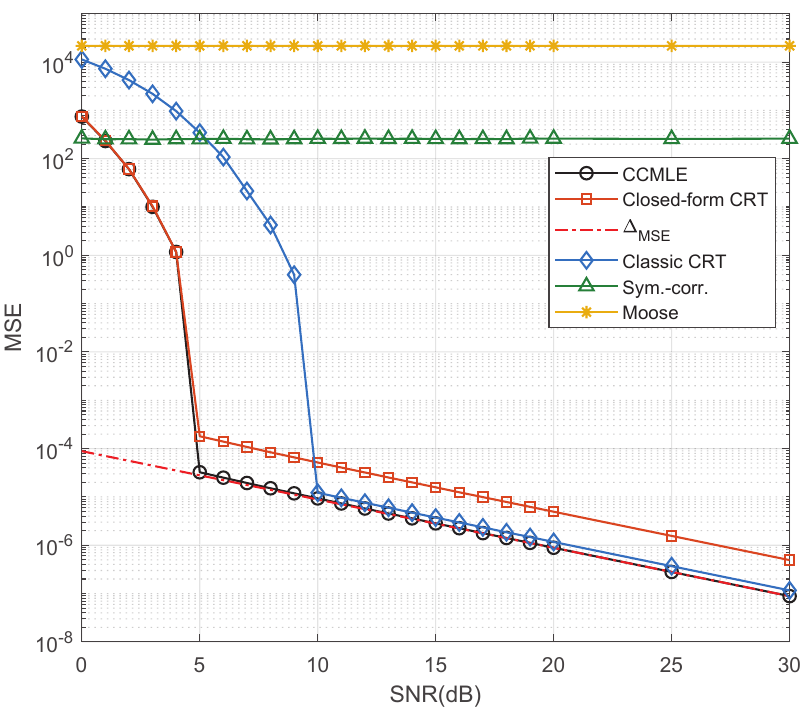}
		\caption{$N=512$}
		\label{N3}
	\end{subfigure}

	\caption{MSE Performance for different parameter configurations}
	\label{sim512}
\end{figure*}

The simulated IER curve v.s. SNR and the theoretical SNR threshold $\eta_{th}$ are presented in Fig. \ref{simTH}.
It can be seen that the theoretical SNR threshold obtained from \textit{Theorem 2} are slightly lower than that from the simulation, which is caused by the difference between the exact distribution of the phase error and the assumed wrapped normal distribution.
In addition, for a better comparison, we present the corresponding results in TABLE \ref{tableTH}, where $\eta_{\rm{sim}}$ represents the SNR threshold obtained from simulations.
It is worth noting that the differences between the theoretical SNR thresholds and the simulated ones are significant at high IER ($\ge 0.001$), but reduce to the range of 0.5 dB to 0.2 dB as IER decreases from $1\times 10^{-3}$ to $1\times 10^{-6}$.
This is due to the fact that the exact distribution of the normalized CFO error, i.e., the von Mises distribution, becomes closer to the wrapped normal distribution with decreasing variance \cite{mardia2000directional}, namely, increasing SNR.
As a result, given a sufficiently small IER, usually less than or equal to 0.001, we can estimate the approximate SNR threshold of the proposed CCMLE in advance based on the parameter configurations without expensive simulations, which enables an efficient waveform design.



\subsection{Parameter Configuration Evaluation}
In this section, we first theoretically analyze the performance for different values of $N$ according to the configuration guidelines in Section-\ref{Guide}.
Next, we carry out simulations to validate the effectiveness of the analysis.
Finally, we outline the impact on performance if the configuration guidelines are not followed.
Specifically, we set the number of subcarriers to be $N=128,256,512$, respectively, and $\varepsilon_N $ is still sampled uniformly from $ [-N/2,N/2]$.

According to the guidelines in Section-\ref{Guide}, subject to the constraint that the maximum sample interval $L_1$ to be less than $N$, we have the parameter configurations shown in TABLE \ref{tableGuide}, where $\eta_{th}$ is obtained by $\delta=1\times 10^{-6}$.
Note that, base on (\ref{eq30a}), we have that $\Delta_{MSE}$ is inversely proportional to $\Sigma_L=\sum\limits_{i=1}^K{L_{i}^3}$ for a given SNR.
Thus, we can use $\Sigma_L$ to evaluate the theoretical MSE.
Next, we further confirm the parameter configurations in the following:
\begin{itemize}[itemsep=0pt,topsep=0pt,parsep=0pt]
    \item{For $N=128$, as both $\Sigma_L$ and $\eta_{th}$ have advantages for the first parameter configuration, we set parameters to be $\Gamma_1=2,\Gamma_2=3,\Gamma_3=5,\Gamma_4=7$.}
    \item{For $N=256$, the theoretical MSE for the second parameter configuration has an order of magnitude increase compared to the first one (about 12 times), but only a 1.5 dB loss in $\eta_{th}$.
Hence, we prefer to set parameters to be $\Gamma_1=11,\Gamma_2=13,\Gamma_3=17$.}
    \item{For $N=512$, the $\eta_{th}$ for the first parameter configuration has a 2.4 dB advantage compared to the second one, but the degradation of theoretical MSE is far less than an order of magnitude (about 2 times).
Hence, we prefer to set parameters to be $\Gamma_1=3,\Gamma_2=5,\Gamma_3=7,\Gamma_4=11$.}
\end{itemize}

\begin{table}[!htbp]
\caption{Parameter configurations with corresponding performance analysis for $N=128,256,512$}
\tabcolsep=0.15cm
\renewcommand\arraystretch{2}
\begin{center}
\begin{tabular}{c|c|c|c|c}
  \hline
                    & $\Gamma_{i}$ &${L_{i}}$ & $\Sigma_L$ & \makecell[l]{$\eta_{th}$ \\(dB)}\\ \hline
  \multirow{2}{*}{128}  & \makecell[c]{$\Gamma_1=2,\Gamma_2=3$\\ $\Gamma_3=5,\Gamma_4=7$} & \makecell[c]{$L_1=105,L_2=70$\\ $L_3=42,L_4=30$} & 1.6$\times$$10^6$ & 6.1 \\ \cline{2-5}
                        & \makecell[c]{$\Gamma_1=5,\Gamma_2=7$\\ $\Gamma_3=11$} & \makecell[c]{$L_1=77,L_2=55$\\ $L_3=35$} & 6.7$\times$$10^5$ & 9.5 \\ \hline\hline
  \multirow{2}{*}{256}  & \makecell[c]{$\Gamma_1=2,\Gamma_2=3$\\ $\Gamma_3=5,\Gamma_4=7$} & \makecell[c]{$L_1=105,L_2=70$\\ $L_3=42,L_4=30$} & 1.6$\times$$10^6$ & 6.1 \\ \cline{2-5}
                        & \makecell[c]{$\Gamma_1=11,\Gamma_2=13$\\ $\Gamma_3=17$} & \makecell[c]{$L_1=221,L_2=187$\\ $L_3=143$} & 2.0$\times$$10^7$ & 7.6 \\ \hline\hline
  \multirow{2}{*}{512}  & \makecell[c]{$\Gamma_1=3,\Gamma_2=5$\\ $\Gamma_3=7,\Gamma_4=11$} & \makecell[c]{$L_1=385,L_2=231$\\ $L_3=165,L_4=105$} & 7.5$\times$$10^7$ & 4.5 \\ \cline{2-5}
                        & \makecell[c]{$\Gamma_1=17,\Gamma_2=19$\\ $\Gamma_3=23$} & \makecell[c]{$L_1=437,L_2=391$\\ $L_3=323$} & 1.8$\times$$10^8$ & 6.9 \\ \hline
\end{tabular}
\end{center}\label{tableGuide}
\end{table}

\par The corresponding simulation results of MSE performances for different parameter configurations are shown in Fig. \ref{sim512}.
It can be seen that the SNR threshold for $N=128,256,512$ falls into the range (6, 7), (7, 8), (4, 5) dB, respectively, which is consistent with the results that $\eta_{th}=6.1, 7.6, 4.5$ dB in TABLE \ref{tableGuide}.

In addition, to illustrate the impact of not following the configuration guidelines in Section-\ref{Guide}, we carry out experiments with an alternative parameter configuration for $N=512$.
Specifically, we set $\Gamma_1=2$, $\Gamma_2=5$, $\Gamma_3=7$, $\Gamma_4=13$, corresponding to $L_1=455$, $L_2=182$, $L_3=130$, $L_4=70$.
In such case, $\Gamma_i(i=1,\ldots,4)$ is no longer $4$ consecutive co-prime numbers, but the sample intervals $L_i$ are still large enough, where $\Sigma_L=1.0\times10^8$ is slightly larger than $\Sigma_L=7.5\times10^7$.
Besides, according to \textit{Theorem 2}, we can obtain $\eta_{th} = 7.7 \,\rm{dB}$, which is 3.2 dB higher than 4.5 dB.

\begin{figure}[!t]
\centerline{\includegraphics[width=3.5 in]{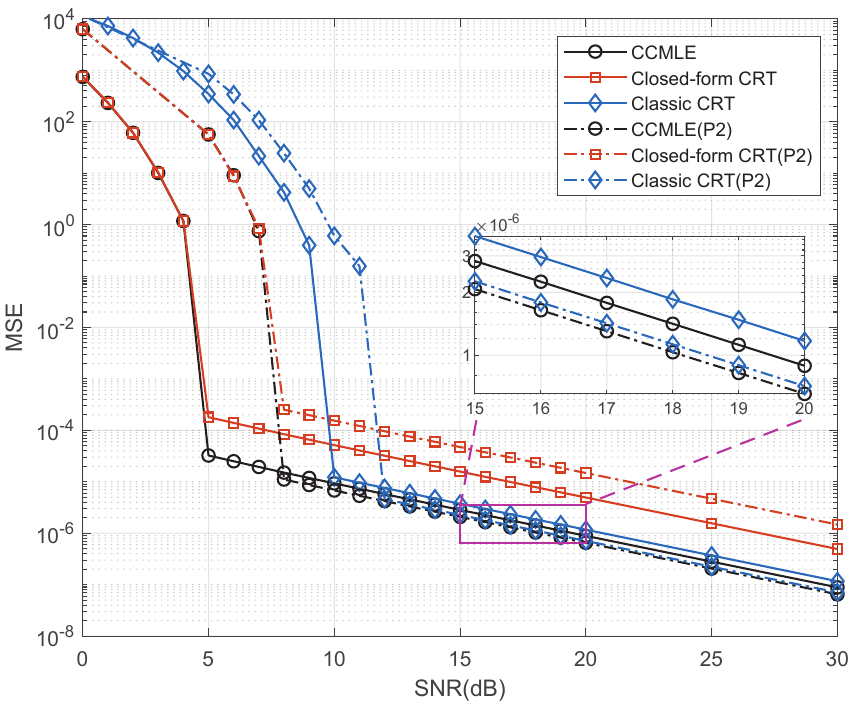}}
\caption{MSE performance for two different group of parameter configurations for the CCMLE method ($N=512$).}
\label{sim512-2}
\end{figure}

Accordingly, the experimental results are presented in Fig. \ref{sim512-2}, where the results for the alternative parameter configuration are marked with dash-dot line and ``(P2)''.
It can be seen that the MSE performance of the CCMLE is similar for two parameter configurations in the absence of IFO errors, i.e., $\rm{SNR}\ge 8\,\rm{dB}$, due to the fact that $\Sigma_L$ is guaranteed to be of approximately the same magnitude in both cases.
However, the experimental results show that the CCMLE loses 3 dB of SNR threshold if the parameters are not configured according to the configuration guidelines in Section-\ref{Guide}, which is consistent with the above analysis.
Moreover, it is worth noting that there are two other significant differences.\\
\indent
First, there is a significant increase in the performance gap between the CCMLE and the Closed-form CRT in Fig. \ref{sim512-2},
which is due to the fact that the differences among $L_i^3$ become more significant for the alternative parameter configuration.
According to (\ref{eq11}), we know that the more significant differences among $L_i^3$ will lead to more significant differences in variance for different sample intervals.
Therefore, the case of the alternative parameter configuration deviates farther from the assumption of equal variance in the Closed-form CRT \cite{wang2015maximum}, which results in the MSE performance further away from the CCMLE.\\
\indent
Second, the achievable performances for the CCMLE and the Classic CRT are much closer,
which is due to the fact that the variance of the maximum sample interval is far smaller than that of the other sample intervals.
Based on (\ref{eq11}) and (\ref{eq19}), it can be found that $\Delta_{MSE(M)}$ and $\sigma_1^2$ will be very close to each other under the condition that $L_1^3$ is much greater than $L_i^3\,(i=2,\ldots,K)$, i.e., the variance of the maximum sample interval is far smaller than that of the other sample intervals.
Therefore, the CCMLE and the Classic CRT can achieve similar performance in high-SNR range, i.e., $\rm{SNR}\ge 12\,\rm{dB}$.
However, the enlarged diagram in Fig. \ref{sim512-2} shows that the CCMLE still slightly outperforms the Classic CRT for $\rm{SNR}\ge 12\,\rm{dB}$, which is consistent with the \textit{Theorem 1}.

\section{Conclusion}\label{Conclusion}
We have proposed the CCMLE method for joint integer and fractional CFO estimation for high-mobility OFDM systems.
This approach enabled a straightforward calculation of the CFO by utilizing various estimates from different ranges without adding significant complexity.
The theoretical analyses for the proposed method were presented, including the MSE performance and the SNR threshold approaching the CRB.
Furthermore, based on the results of theoretical analysis, we presented the guideline for selecting the parameter configuration to accommodate a better MSE performance and a lower SNR threshold.
Finally, extensive experiments demonstrated the consistency with the theoretical analysis and showed that the proposed CCMLE method offers better performance than the baseline schemes.
\appendices
\section{\textit{Proof of Theorem 1}}\label{appendixT1}
\setcounter{equation}{0}
 \renewcommand{\theequation}{A.\arabic{equation}}
\textit{Proof:}
By substituting the weights $w_i$ defined in (\ref{eq15}) into (\ref{eq16}), we can simplify (\ref{eq16}) as
\begin{equation}
\begin{array}{l}
\displaystyle \Delta_{MSE(M)} = \frac{{1}}{{\sum\limits_{i = 1}^K \frac{1}{\sigma _i^2} }}.
\end{array}\label{eq19}
\end{equation}
Since $\sigma^2_i>0$ for $i=1,\ldots, K$, we have
\begin{equation}
\begin{array}{l}
\displaystyle {{\sum\limits_{i = 1}^K \frac{1}{\sigma _i^2} }} > \frac{1}{\sigma _j^2},\,j=1,\ldots,K.
\end{array}\label{eq20}
\end{equation}

\noindent Clearly, we have
\begin{equation}
\begin{array}{l}
\displaystyle \Delta_{MSE(M)} < \sigma _j^2,\,j=1,\ldots,K,
\end{array}\label{eq17}
\end{equation}
which indicates that $\Delta_{MSE(M)}$ is smaller than the minimum among $\sigma^2_i$ ($i=1,\ldots, K$), i.e., $\Delta_{MSE(M)} < \text{min}\{\sigma^2_i\}$.\par

Next, substitute $\Gamma,\,\varepsilon$ by $M\Gamma$, $\varepsilon_M$ in (\ref{eq7}) and plug it into (\ref{eq8}), we can obtain the following sample vector:
\begin{equation}
\begin{array}{l}
{\bf{Z}} = \left[ Z_1,...,Z_K \right],
\end{array}\label{eq24}
\end{equation}
where
\begin{equation}
\begin{array}{l}
Z_i= L_i{e^{j \frac{2\pi L_i}{M\Gamma}\varepsilon_M}} + {\omega_{L_i}},
\end{array}\label{eq25}
\end{equation}
and $\omega_{L_i}$ is the complex Gaussian noise sample with mean zero and variance $2{L_i} \sigma^2$.
Note that $\sigma^2$ represents the noise variance of the received signal $r(n)$.\\
\indent According to \cite{rife1974single}, the joint PDF of the elements of the sample vector ${\bf{Z}}$ is given by
\begin{equation}
\begin{array}{l}
f({\bf{Z}};{\varepsilon _M}) \vspace{1ex}\\
= {\left( {\frac{1}{{2\pi }}} \right)^K}\prod\limits_{i = 1}^K {\frac{1}{{\sigma _{{L_i}}^2}}} \exp \left\{ { - \sum\limits_{i = 1}^K {\frac{{{{\left( {{X_i} - {\mu _i}} \right)}^2} + {{\left( {{Y_i} - {\nu _i}} \right)}^2}}}{{2\sigma _{{L_i}}^2}}} } \right\},
\end{array}\label{eq26}
\end{equation}
where $X_i$ and $Y_i$ represent the real and imaginary parts of the sample $Z_i$, respectively. $\sigma _{{L_i}}^2=L_i \sigma^2$ represents the noise variance of $X_i$ and $Y_i$, and
\begin{equation}
\begin{array}{l}
\displaystyle{\mu _i} = {L_i}\cos \left( {\frac{2\pi L_i}{M\Gamma}\varepsilon_M} \right),\vspace{1ex}\\
\displaystyle{\nu _i} = {L_i}\sin \left( {\frac{2\pi L_i}{M\Gamma}\varepsilon_M} \right).
\end{array}\label{eq27}
\end{equation}

\noindent Then, the elements of the Fisher Information Matrix (FIM) $\bf{J}$ can be derived as follows \cite{rife1974single}:
\begin{equation}
\begin{array}{l}
\displaystyle{J_{mn}} = \sum\limits_{i = 1}^K {\frac{1}{{\sigma _{{L_i}}^2}}\left[ {\frac{{\partial {\mu _i}}}{{\partial {\alpha _m}}}\frac{{\partial {\mu _i}}}{{\partial {\alpha _n}}} + \frac{{\partial {\nu _i}}}{{\partial {\alpha _m}}}\frac{{\partial {\nu _i}}}{{\partial {\alpha _n}}}} \right]},
\end{array}\label{eq28}
\end{equation}
where $\alpha_m (m=1,2,\ldots)$ are unknown parameters.

\indent {Note that the only parameter to be estimated in (\ref{eq24}) is $\varepsilon_M$, i.e., $\alpha_1=\varepsilon_M$.}
Therefore, the CRB of $\varepsilon_M$ is given by \cite{rife1974single}
\begin{equation}
\begin{aligned}
\displaystyle
var[\hat\varepsilon_{M}] \ge J_{11}^{-1} = \left({\sum\limits_{i = 1}^K \frac{{{4{\pi ^2}L_i^3}}}{{\sigma ^2 M^2\Gamma ^2}}}\right)^{-1}.
\end{aligned}\label{eq29}
\end{equation}
Specifically, from (\ref{eq2n}) and (\ref{eq1}), we know that the SNR of the received signals can be represented as $\eta=1/\sigma ^2$.
Hence, based on (\ref{eq11}) and (\ref{eq19}), we have
\begin{equation}
\begin{aligned}
\displaystyle
var[\hat\varepsilon_{M}] &\ge \frac{1}{\sum\limits_{i = 1}^K \frac{{{4{\pi ^2}L_i^3 \cdot \eta}}}{{M^2\Gamma ^2}}}
=\frac{{1}}{{\sum\limits_{i = 1}^K \frac{1}{\sigma _i^2} }}.
\end{aligned}\label{eq30}
\end{equation}
\par
\noindent This proves the theorem.$\hfill\blacksquare$

\section{\textit{Proof of Lemma 1}}\label{appendixA}
\setcounter{equation}{0}
 \renewcommand{\theequation}{B.\arabic{equation}}

\textit{Proof:}
\begin{itemize}
\item{\textit{Case I:} $S \ne \emptyset ,\bar S \ne \emptyset$}
\end{itemize}

Let
\begin{equation}\label{A1}
f(x)=\frac{1}{x}+\frac{1}{a-x},\,0<x<a,
\end{equation}
where $a$ is a constant. Then, we have the derivative of $f(x)$ as
\begin{equation}\label{A1a}
f'(x)=-\frac{1}{x^2}+\frac{1}{(a-x)^2},\,0<x<a.
\end{equation}
Clearly, we have
\begin{equation}\label{A1b}
\begin{aligned}
f'(x)>0,\quad \text{if}\quad \frac{a}{2}<x<a,\\
f'(x)<0,\quad \text{if}\quad 0<x<\frac{a}{2}.
\end{aligned}
\end{equation}
Therefore, $f(x)$ is monotonically decreasing on $(0,\frac{a}{2})$ and is monotonically increasing on $(\frac{a}{2},a)$.

\noindent Let
\begin{equation}\label{A2xx}
P = \left\{\sum\limits_{{{\Delta \varepsilon }_{{M_i}}} \in S} {L_i^3}\right\}.
\end{equation}

\noindent Based on the fact that $S$ is the subset of $U$, and $\bar S$ is the complement of $S$ in $U$, let
\begin{equation}\label{A3}
\begin{aligned}
a-x &= \sum\limits_{{{\Delta \varepsilon }_{{M_j}}} \in \bar S} {L_j^3}.
\end{aligned}
\end{equation}
where $ x \in P$ and
\begin{equation}\label{A2b}
\begin{aligned}
a=\sum\limits_{i=1}^K {L_i^3}.
\end{aligned}
\end{equation}


\noindent As a result, we have
\begin{equation}\label{A4}
\xi_\Psi=f(x), x \in P,
\end{equation}
where the constant $a$ is defined by (\ref{A2b}).

\noindent Note that
\begin{equation}\label{A7}
L_1 > \cdots > L_K>0.
\end{equation}

\noindent Hence, according to (\ref{A2xx}), (\ref{A2b}) and (\ref{A7}), we have
\begin{equation}\label{A8}
\begin{aligned}
x_{\rm{min}} &= L_K^3<\frac{a}{2},\\
x_{\rm{max}} &= \sum\limits_{i=1}^{K-1} {L_i^3}>\frac{a}{2}.
\end{aligned}
\end{equation}
Recall that $f(x)$ is monotonically decreasing on $(0,\frac{a}{2})$ and is monotonically increasing on $(\frac{a}{2},a)$.
Thus, we have
\begin{equation}\label{A9}
\begin{aligned}
f(x_{\rm{min}}) &\ge f(x),\quad \text{if}\quad x<\frac{a}{2},\\
f(x_{\rm{max}}) &\ge f(x),\quad \text{if}\quad x>\frac{a}{2}
\end{aligned}
\end{equation}
for $x \in P$.

Clearly, for $x_1,x_2 \in P$, if $x_1 + x_2 = a$, we have
\begin{equation}\label{A10}
f(x_1)=f(x_2).
\end{equation}

\noindent According to (\ref{A8}), we have $x_{\rm{min}} + x_{\rm{max}} = a$. Hence,
\begin{equation}\label{A11}
f(x_{\rm{min}})=f(x_{\rm{max}}).
\end{equation}
Let
\begin{equation}\label{A11ss}
\xi_\Psi^*=\rm{max}\{\xi_\Psi\}.
\end{equation}
Based on (\ref{A4}), (\ref{A9}) and (\ref{A11}), we have that $\xi_\Psi^*$ is the maximum value of $f(x)$, given by
\begin{equation}\label{A12aa}
\xi_\Psi^*=f(x_{\rm{min}})=f(x_{\rm{max}}).
\end{equation}
That is,
\begin{equation}\label{A12}
\xi_\Psi^*={\frac{1}{L_K^3}} + {\frac{1}{{\sum\limits_{j\ne K} {L_j^3} }}}.
\end{equation}

\begin{itemize}
\item{\textit{Case II:} $S = \emptyset {\rm{\, or \,}}\bar S = \emptyset$}
\end{itemize}

If $S$ or $\bar S$ is an empty set, according to (\ref{eq48}) and (\ref{A7}), we have
\begin{equation}\label{A13}
\begin{aligned}
\xi_\Psi={\frac{1}{\sum\limits_{i=1}^K {L_i^3}}} &< \frac{1}{L_K^3}<\xi_\Psi^*.
\end{aligned}
\end{equation}

\noindent This proves the lemma.$\hfill\blacksquare$

\section{\textit{Proof of Theorem 2}}\label{appendixT2}

\setcounter{equation}{0}
 \renewcommand{\theequation}{C.\arabic{equation}}
\textit{Proof:}
According to \cite{wang2015maximum}, we have that the normalized CFO estimation error of CCMLE methods $\Delta\varepsilon_{M}=\sum\limits_{i = 1}^K {w_i} \Delta\varepsilon_{M_i}$ holds, i.e., the MSE $\Delta_{MSE(M)} = \sum\limits_{i = 1}^K {w_i^2} \sigma _i^2$ approaches the CRB holds, if and only if
\begin{equation}
\begin{array}{l}
\left| \Psi  \right| < {M \mathord{\left/{\vphantom {M 2}} \right. \kern-\nulldelimiterspace} 2}
\end{array}\label{eq38}
\end{equation}
holds for any subset $S$ of set $U=\left\{{{\Delta \varepsilon }_{{M_1}}}, \ldots,{{\Delta \varepsilon }_{{M_K}}}\right\}$,
where $\Psi$ is defined as \cite{wang2015maximum}
\begin{equation}
\displaystyle
\Psi=\sum\limits_{{{\Delta \varepsilon }_{{M_i}}} \in S} {\frac{{{w_i}{{\Delta \varepsilon }_{{M_i}}}}}{{\sum\nolimits_{{{\Delta \varepsilon }_{{M_j}}} \in S} {{w_j}} }}}  - \sum\limits_{{{\Delta \varepsilon }_{{M_i}}} \in \bar S} {\frac{{{w_i}{{\Delta \varepsilon }_{{M_i}}}}}{{\sum\nolimits_{{{\Delta \varepsilon }_{{M_j}}} \in \bar S} {{w_j}} }}}
\label{eq39}
\end{equation}
and $\bar S$ is the complement of $S$ in $U$.
Therefore, $\Delta_{MSE(M)}$ approaching the CRB with a probability of at least $1-\delta$ means $$p\left( {\left| \Psi \right| < {M \mathord{\left/{\vphantom {M 2}} \right.\kern-\nulldelimiterspace} 2}} \right) > 1- \delta,$$ i.e., $p\left( {\left| \Psi \right| \ge {M \mathord{\left/{\vphantom {M 2}} \right.\kern-\nulldelimiterspace} 2}} \right) < \delta$.

Furthermore, according to (\ref{eq39}), since $\Psi$ is the linear combination  of ${\Delta \varepsilon }_{M_i}$ that follows the wrapped normal distribution, $\Psi$ also follows the wrapped normal distribution \cite{jammalamadaka2001topics}.
Hence, the probability $p\left( {\left| \Psi \right| \ge {M \mathord{\left/{\vphantom {M 2}} \right.\kern-\nulldelimiterspace} 2}} \right)$ can be obtained by integrating the PDF of $\Psi$, i.e.,
\begin{equation}
\displaystyle
p\left( {\left| \Psi  \right| \ge {M \mathord{\left/{\vphantom {M 2}} \right.\kern-\nulldelimiterspace} 2}} \right)= 2\int_{ {M \mathord{\left/
 {\vphantom {M 2}} \right.
 \kern-\nulldelimiterspace} 2}}^{A} {{f_\Psi }(x)dx},
\label{eq40}
\end{equation}
where $A \le M\Gamma_1$ is the maximum of the wrapped normal distributed random variables. ${{f_\Psi }(x)}$ is the PDF of the wrapped normal distribution with mean zero and variance $\sigma_{\Psi}^2$, represented as \cite{jammalamadaka2001topics}
\begin{equation}
{f_\Psi }(x) = \frac{1}{{{\sigma _\Psi }\sqrt {2\pi } }}\sum\limits_{k =  - \infty }^{ + \infty } {\exp \left\{ { - \frac{{{{(x + 2kA)}^2}}}{{2\sigma _\Psi ^2}}} \right\}},
\label{eq41}
\end{equation}
{where $-A \le x <A$ and $k$ is an integer.}

Let $t=x + 2kA$, we can rewrite (\ref{eq40}) as the integral of the normal distribution over the interval $\Sigma$, as follows:
\begin{equation}
\begin{array}{l}
p\left( {\left| \Psi  \right| \ge {M \mathord{\left/{\vphantom {M 2}} \right.\kern-\nulldelimiterspace} 2}} \right) \vspace{1.5ex}\\
\displaystyle= \frac{2}{{{\sigma _\Psi }\sqrt {2\pi } }}\int_{ {M \mathord{\left/
 {\vphantom {M 2}} \right.
 \kern-\nulldelimiterspace} 2}}^{A} {\sum\limits_{k =  - \infty }^{ + \infty } {\exp \left\{ { - \frac{{{{(x + 2kA)}^2}}}{{2\sigma _\Psi ^2}}} \right\}}dx}\vspace{1.5ex}\\
\displaystyle=\frac{2}{{{\sigma _\Psi }\sqrt {2\pi } }}\int_{ \Sigma } { {\exp \left( { - \frac{{{t^2}}}{{2\sigma _\Psi ^2}}} \right)}dt}.
\label{eq42}
\end{array}
\end{equation}
Based on the symmetry of normal distribution, $\Sigma$ can be further expressed as
\begin{equation}\label{eq44}
\Sigma =\left[\frac{M}{2},2A\!-\!\frac{M}{2}\right]\cup\left[\frac{M}{2}\!+\!2A,4A\!-\!\frac{M}{2}\right]\cup\cdots.
\end{equation}

\noindent Therefore, the probability $p\left( {\left| \Psi \right| \ge {M \mathord{\left/{\vphantom {M 2}} \right.\kern-\nulldelimiterspace} 2}} \right)$ in (\ref{eq42}) can be simplified as
\begin{equation}
\begin{array}{l}
p\left( {\left| \Psi  \right| \ge {M \mathord{\left/{\vphantom {M 2}} \right.\kern-\nulldelimiterspace} 2}} \right) \vspace{1ex}\\
\displaystyle=
\displaystyle 2 Q\left( {\frac{M}{{2{\sigma _\Psi }}}} \right) - 2 Q\left( {\frac{{2A}}{{{\sigma _\Psi }}} - \frac{M}{{2{\sigma _\Psi }}}} \right) + \vspace{1.5ex}\\
\displaystyle \quad 2 Q\left( {\frac{{2A}}{{{\sigma _\Psi }}} \!+\! \frac{M}{{2{\sigma _\Psi }}}} \right) - 2 Q\left( {\frac{{4A}}{{{\sigma _\Psi }}} \!-\! \frac{M}{{2{\sigma _\Psi }}}} \right) +  \cdots
\vspace{1ex}\\
\displaystyle \approx 2Q\left(\frac{M}{2\sigma _\Psi}\right),
\label{eq45}
\end{array}
\end{equation}
where $Q(t)$ is $Q$-function defined by
\begin{equation}\label{eq46}
Q(t)=\frac{1}{{\sqrt {2\pi } }}\int_{ t }^{+ \infty} { {\exp \left( { - \frac{{{t^2}}}{{2}}} \right)}dt}.
\end{equation}

\noindent According to (\ref{eq11}) and (\ref{eq15}), the variance of $\Psi$ can be represented as a function with respect to SNR, as follows:
\begin{equation}
\begin{aligned}
\displaystyle \sigma_{\Psi}^2(\eta)&={\frac{1}{{\sum\limits_{{{\Delta \varepsilon }_{{M_i}}} \in S} {\frac{1}{\sigma _i^2}} }}} + {\frac{1}{{\sum\limits_{{{\Delta \varepsilon }_{{M_j}}} \in \bar S} {\frac{1}{\sigma _j^2}} }}}\\
&=\frac{M^2\Gamma^2}{4\pi^2\eta} \xi_\Psi,
\label{eq47}
\end{aligned}
\end{equation}
where $\xi_\Psi$ is given by (\ref{eq48}).

Recall that $\left| \Psi  \right| < {M \mathord{\left/{\vphantom {M 2}} \right. \kern-\nulldelimiterspace} 2}$ holds for any subset $S$ of set $U$, which means that $p\left( {\left| \Psi \right| \ge {M \mathord{\left/{\vphantom {M 2}} \right.\kern-\nulldelimiterspace} 2}} \right) < \delta$ holds for any $\sigma_{\Psi}^2(\eta)$ under a given $\eta$.
In addition, note that $p\left( {\left| \Psi \right| \ge {M \mathord{\left/{\vphantom {M 2}} \right.\kern-\nulldelimiterspace} 2}} \right)$ increases as $\sigma_{\Psi}^2(\eta)$ increases.

Thus, the above condition can be equivalently converted to $p\left( {\left| \Psi \right| \ge {M \mathord{\left/{\vphantom {M 2}} \right.\kern-\nulldelimiterspace} 2}} \right) < \delta$ holds for $\rm{max}\{\sigma_{\Psi}^2(\eta)\}$ under a given $\eta$, where $\rm{max}\{\sigma_{\Psi}^2(\eta)\}$ represents the maximum of ${\sigma_{\Psi}^2(\eta)}$.

Meanwhile, based on (\ref{eq47}) and (\ref{eq48}), for a given $\eta$, the maximum of ${\sigma_{\Psi}^2(\eta)}$ is given by
\begin{equation}\label{eq49}
\rm{max}\{\sigma_{\Psi}^2(\eta)\}=\frac{M^2\Gamma^2}{4\pi^2\eta} \rm{max}\{\xi_\Psi\},
\end{equation}
where $\rm{max}\{\xi_\Psi\}$ represents the maximum value of $\xi_\Psi$.
As a result, the problem of finding the maximum value of the variance $\sigma_{\Psi}^2(\eta)$ is simplified to the problem of determining the maximum value of $\xi_\Psi$ defined by (\ref{eq48}).

Let $\rm{max}\{\xi_\Psi\}=\xi_\Psi^*$, according to \textit{Lemma 1} in Appendix \ref{appendixA}, we have that $\xi_\Psi^*$ is given by (\ref{A12}).

Let
\begin{equation}\label{eq50}
\displaystyle \delta=2Q\left(\frac{M}{2\sigma^{\rm{max}} _\Psi(\eta)}\right)=2Q\left(x_\delta\right),
\end{equation}
where
\begin{equation}\label{eq51}
\displaystyle \sigma^{\rm{max}} _\Psi(\eta)=\sqrt{\rm{max}\{\sigma_{\Psi}^2(\eta)\}}.
\end{equation}

\noindent
According to the monotonicity of the $Q$-function, we have
\begin{equation}\label{eq52}
\frac{M}{2\sigma^{\rm{max}} _\Psi(\eta)}=x_\delta,
\end{equation}
where the value of $x_\delta$ can be obtained by solving (\ref{eq50}) given an arbitrarily small number $\delta$.
In addition, some typical approximate values of $x_\delta$ corresponding to $\delta$ are presented in TABLE \ref{table0}.
\begin{table}[!htbp]
\caption{The approximate relationship between $x_\delta$ and $\delta$}
\renewcommand\arraystretch{1.6}
\begin{center}
\begin{tabular}{c|c|c|c}
  \hline
  $\delta$   & $1\times 10^{-1}$ & $1\times 10^{-2}$ & $1\times 10^{-3}$\\ \hline
  $x_\delta$ & $\approx$ 1.7 & $\approx$ 2.6 & $\approx$ 3.3\\ \hline
  $\delta$   & $1\times 10^{-4}$ & $1\times 10^{-5}$ & $1\times 10^{-6}$\\ \hline
  $x_\delta$ & $\approx$ 3.9 & $\approx$ 4.4 & $\approx$ 4.9 \\
  \hline
\end{tabular}
\end{center}\label{table0}
\end{table}

\noindent Furthermore, the variance of $\Psi$ decreases as the SNR increases, which leads to the integral result of (\ref{eq40}) being even smaller.
Therefore, based on (\ref{eq45}), (\ref{eq49}), (\ref{eq51}) and (\ref{eq52}), we can conclude that $p\left( {\left| \Psi \right| \ge {M \mathord{\left/{\vphantom {M 2}} \right.\kern-\nulldelimiterspace} 2}} \right) < \delta$ holds for 
\begin{equation}\label{eq53xx}
\eta \ge \frac{\Gamma^2 x_\delta^2 \xi_\Psi^*}{\pi^2},
\end{equation}
where $\xi_\Psi^*$ is given by (\ref{A12}) and $x_\delta$ can be obtained by solving (\ref{eq50}) given an arbitrarily small number $\delta$.

Let
\begin{equation}\label{eq53}
\eta_{th} = \frac{\Gamma^2 x_\delta^2 \xi_\Psi^*}{\pi^2}.
\end{equation}
\noindent Then, we have $$p\left( {\left| \Psi \right| < {M \mathord{\left/{\vphantom {M 2}} \right.\kern-\nulldelimiterspace} 2}} \right) > 1- \delta$$ for $\eta \ge \eta_{th}$, i.e., the MSE of the proposed CCMLE method approachs the CRB with a probability of at least $1-\delta$ for $\eta \ge \eta_{th}$, where a closed-form $\eta_{th}$ is given by (\ref{eq53}).
%
%
This proves the theorem.$\hfill\blacksquare$

\bibliography{Refer_MLE_CRT_CFO}
\bibliographystyle{IEEEtran}

\end{document}